# Removing redundancy in relativistic quantum mechanics

## Peter Rowlands

*Department of Physics, University of Liverpool, Oliver Lodge Laboratory, Oxford Street, Liverpool, L69 7ZE, UK. e-mail p.rowlands@liverpool.ac.uk*

*Abstract* It is proposed that the Dirac equation, as normally interpreted, incorporates intrinsic redundancies whose removal necessarily leads to an enormous gain in calculating power and physical interpretation. Streamlined versions of the Dirac equation can be developed which remove the redundancies and singularities from many areas of quantum physics while giving quantum representations to specific particle states.

**1 Singularities and redundancy**

The transformation of coordinate systems is a powerful device used in many areas of physics, but its most significant effects occur when the coordinate systems before and after the transformation are not absolutely equivalent. The particular choice of coordinate system will then often determine the subsequent development of the mathematical structure as a physical representation of the system being investigated; and whether or not the transformation is desirable will be determined by how 'correct' this physical representation is considered to be. Spherical polar coordinates, for example, are ideal for systems with point or radial sources – as with the hydrogen atom or the Schwarzschild solution in GR – and this is because they privilege one spatial dimension (the radial) over the others, as required. In these cases the source of the field can be considered as a singularity, but in other cases the choice of such a coordinate system can lead to a singularity appearing where none exists in reality. If we find that the mere choice of a coordinate system leads to singularities, then we have, of course, introduced a problem that needn't exist. But it isn't always only a single problem. Singularities can often be avoided or overcome by using special mathematical techniques or careful definition of the valid 'physical' limits of the system being investigated, but, sometimes, the singularity acts as a barrier, which separates the system into two seemingly unconnected halves, leading to immediate duplication of the information required; in addition, the existence of an uncrossable boundary can lead to *repeated* duplication of information as the system attempts to compensate in one way or another for the loss of a connectedness that ought to exist.

The choice of coordinate transformations to represent rotation has always been a particularly difficult problem, especially when matrix methods, such as Euler angle rotation in *SO*(3), are used. The origin of the problem is a singularity at the point where the angle $\beta = \pi / 2$. As one author writes: 'Inherent in every minimal Euler angle rotation sequence in *SO*(3) – the group whose elements are the Special Orthogonal matrices in $R^3$ – is at least one singularity.'[1] Essentially, the problem arises from the use of an intrinsically 2-dimensional mathematical structure to represent a 3-dimensional reality; we can, for example, show that the problem is immediately solved and the singularity removed when the intrinsically 3-dimensional quaternions are introduced.[1] Exactly, the same kind of reasoning can be applied to relativistic quantum mechanics, where problems emerge from the imposition of a matrix representation. Relativistic quantum mechanics as represented by the Dirac equation and quantum field theory produces at least one type of singularity that appears to be an artefact of the system – the infrared divergence. It also leads to infinities that have to be removed by renormalization, even in the ideal case of free particles where there is apparently no real source for the divergent terms. In addition, there appears to be a great deal of redundancy. For example, QCD calculations using Feynman diagrams derived from the standard gamma matrix representation require ten million calculations for a six gluon interaction, whereas the alternative algebraic approach using twistor space, originally proposed by Witten,[2] reduces the calculations required to only six. Even with this method, it is clear that redundancies are still visible; so the question we should ask is whether it is possible to find a coordinate system for the fermionic state which removes redundancy entirely.

**2 Redundancy in the Dirac equation**

The Dirac equation, as conventionally written, in matrix form,
$$(\gamma^\mu \partial_\mu + im)\, \psi = 0 \ , \qquad (1)$$
though apparently compact, in fact contains a large amount of redundancy. The ultimate source of this redundancy is a faultline in the matrix representation for the gamma operators which is most clearly manifested in the three momentum operators. Here, we find a system which is not rotation symmetric, unlike physical momentum. The mathematical constraints brought about by using matrices force us into a physical representation which does not reflect reality, and which may therefore be inadvertently introducing a 'redundancy barrier' of the kind discussed in the previous section. (In fact, any mathematical representation which makes a 3-D system rotation asymmetric may be considered either the origin or the signature of a singularity.)

The gamma operators in matrix form are normally expressed using Pauli matrices. To understand the more fundamental algebra involved, however, we need first to look at quaternions. Quaternions (represented by bold italics) are algebraic operators multiplying according to the rules:
$$\boldsymbol{i}^2 = \boldsymbol{j}^2 = \boldsymbol{k}^2 = \boldsymbol{ijk} = -1$$
$$\boldsymbol{ij} = -\boldsymbol{ji} = \boldsymbol{k}$$
$$\boldsymbol{jk} = -\boldsymbol{kj} = \boldsymbol{i}$$
$$\boldsymbol{ki} = -\boldsymbol{ik} = \boldsymbol{j} \ . \qquad (2)$$

If we now complexify the quaternions (i.e. multiply by pseudoscalar $i$), we obtain the products:
$$(ii)^2 = (ij)^2 = (ik)^2 = -i(ii)(ij)(ik) = 1$$
$$(ii)(ij) = -(ij)(ii) = i(ik)$$
$$(ij)(ik) = -(ik)(ij) = i(ii)$$
$$(ik)(ii) = -(ii)(ik) = i(ij).$$

Another way of writing these rules is to use *multivariate vectors* (represented by bold characters):
$$\mathbf{i}^2 = \mathbf{j}^2 = \mathbf{k}^2 = -i\mathbf{ijk} = 1$$
$$\mathbf{ij} = -\mathbf{ji} = i\mathbf{k}$$
$$\mathbf{jk} = -\mathbf{kj} = i\mathbf{i}$$
$$\mathbf{ki} = -\mathbf{ik} = i\mathbf{j}, \qquad (3)$$

where
$$\mathbf{i} = ii, \quad \mathbf{j} = ij, \quad \mathbf{k} = ik,$$

are the units of a vector system whose general 'full' product for two vectors **a** and **b** of is of the form
$$\mathbf{ab} = \mathbf{a}.\mathbf{b} + i\mathbf{a} \times \mathbf{b}.$$

The expressions $i\mathbf{i}$, $i\mathbf{j}$, $i\mathbf{k}$, which emerge from the multiplication rules are the well-known unit pseudovectors.

The product rules for **i**, **j**, **k** are, of course, identical to those for Pauli matrices, $\sigma_x$, $\sigma_y$, $\sigma_z$, and Hestenes used this, as long ago as 1966, to derive the origin of spin from the $i\mathbf{a} \times \mathbf{b}$ term in the full product. Among other things, Hestenes and his followers were able to show how, writing the Schrödinger equation in terms of a multivariate, rather than ordinary, vector $\nabla$, automatically generates fermionic half-integral spin.[3-4] So, defining

$$\sigma_x = \begin{pmatrix} 0 & 1 \\ 1 & 0 \end{pmatrix} \quad \sigma_y = \begin{pmatrix} 0 & -i \\ i & 0 \end{pmatrix} \quad \sigma_z = \begin{pmatrix} 1 & 0 \\ 0 & -1 \end{pmatrix} \quad \text{with unit} \quad I = \begin{pmatrix} 1 & 0 \\ 0 & 1 \end{pmatrix}$$

we have:
$$\sigma_x\sigma_y = -\sigma_y\sigma_x = i\sigma_x$$
$$\sigma_y\sigma_z = -\sigma_z\sigma_y = i\sigma_x$$
$$\sigma_z\sigma_x = -\sigma_x\sigma_z = i\sigma_y.$$

But there is a fundamental difference between these two isomorphic systems. Pauli matrices are not symmetric in three dimensions because they are based on a two-dimensional number system – the complex plane. The notable thing about Pauli matrices is the fact that the multiplication rules for their algebraic equivalents involve an extra pseudoscalar term ($i$) in the pseudovector products. This means that, at least one of the three matrices must have complex coefficients, creating an asymmetric relationship between them..

This complexity must also carry over to the gamma matrices, as these are conventionally defined in terms of Pauli matrix components.

$$\gamma = \begin{pmatrix} 0 & \sigma \\ -\sigma & 0 \end{pmatrix} \quad \gamma_1 = \begin{pmatrix} 0 & \sigma_x \\ -\sigma_x & 0 \end{pmatrix} \quad \gamma_2 = \begin{pmatrix} 0 & \sigma_y \\ -\sigma_y & 0 \end{pmatrix} \quad \gamma_3 = \begin{pmatrix} 0 & \sigma_z \\ -\sigma_z & 0 \end{pmatrix}$$

$$\gamma_0 = \begin{pmatrix} I & 0 \\ 0 & -I \end{pmatrix} \quad \gamma_5 = \begin{pmatrix} 0 & -I \\ -I & 0 \end{pmatrix} \quad \gamma_5 = i\gamma_0\gamma_1\gamma_2\gamma_3$$

leading to:

$$\gamma_1 = \begin{pmatrix} 0 & 0 & 0 & 1 \\ 0 & 0 & 1 & 0 \\ 0 & -1 & 0 & 0 \\ -1 & 0 & 0 & 0 \end{pmatrix} \quad \gamma_2 = \begin{pmatrix} 0 & 0 & 0 & -i \\ 0 & 0 & i & 0 \\ 0 & i & 0 & 0 \\ -i & 0 & 0 & 0 \end{pmatrix} \quad \gamma_3 = \begin{pmatrix} 0 & 0 & 1 & 0 \\ 0 & 0 & 0 & -1 \\ -1 & 0 & 0 & 0 \\ 0 & 1 & 0 & 0 \end{pmatrix}$$

$$\gamma_0 = \begin{pmatrix} 1 & 0 & 0 & 0 \\ 0 & 1 & 0 & 0 \\ 0 & 0 & -1 & 0 \\ 0 & 0 & 0 & -1 \end{pmatrix} \quad I = \begin{pmatrix} 1 & 0 & 0 & 0 \\ 0 & 1 & 0 & 0 \\ 0 & 0 & 1 & 0 \\ 0 & 0 & 0 & 1 \end{pmatrix}$$

With these components, the 4 × 4 Dirac differential operator now becomes:

$$\begin{pmatrix} \frac{\partial}{\partial t} + im & 0 & \frac{\partial}{\partial z} & \frac{\partial}{\partial x} - i\frac{\partial}{\partial y} \\ 0 & \frac{\partial}{\partial t} + im & \frac{\partial}{\partial x} + i\frac{\partial}{\partial y} & -\frac{\partial}{\partial z} \\ -\frac{\partial}{\partial z} & -\frac{\partial}{\partial x} + i\frac{\partial}{\partial y} & -\frac{\partial}{\partial t} + im & 0 \\ -\frac{\partial}{\partial x} - i\frac{\partial}{\partial y} & \frac{\partial}{\partial z} & 0 & -\frac{\partial}{\partial t} + im \end{pmatrix}$$

from which we can derive two positive and two negative energy free-particle spinors for $\psi$, with respective phases $exp(-ip.x)$ and $exp(ip.x)$. The positive energy spinors become:

$$\left(\frac{E+m}{2m}\right)^{1/2} \begin{pmatrix} 1 \\ 0 \\ \frac{p_z}{E+m} \\ \frac{p_x + ip_y}{E+m} \end{pmatrix} \quad \text{and} \quad \left(\frac{E+m}{2m}\right)^{1/2} \begin{pmatrix} 0 \\ 1 \\ \frac{p_x - ip_y}{E+m} \\ \frac{-p_z}{E+m} \end{pmatrix}$$

and the negative energy spinors:

$$\left(\frac{E+m}{2m}\right)^{1/2} \begin{pmatrix} \frac{p_z}{E+m} \\ \frac{p_x + ip_y}{E+m} \\ 1 \\ 0 \end{pmatrix} \quad \text{and} \quad \left(\frac{E+m}{2m}\right)^{1/2} \begin{pmatrix} \frac{p_x - ip_y}{E+m} \\ \frac{-p_z}{E+m} \\ 0 \\ 1 \end{pmatrix}$$

Immediately, we note a potential problem: there is no such physical object as $p_x + ip_y$ or $p_x - ip_y$. Obviously, using the complex notation makes $p_x$ orthogonal to $p_y$, so acting as a kind of substitute vector addition, but it can't be extended to create a 3-dimensional operator on an equivalent basis, and it has the unwelcome consequence of making $p_x$ physically different from $p_y$.

## 3 Defragmenting the Dirac equation

In fact, though the Dirac equation, as conventionally written (1), is the fundamental basis of particle physics, it is inconvenient in many ways in addition to the unphysical nature of the spinor solutions, with main problems relating to its asymmetric structure. Thus the operator ($\gamma^\mu \partial_\mu + im$) is a 4 × 4 matrix, as are each of the four terms $\gamma^\mu$, while $\psi$ is a 4-component (vector) spinor. There are many possible choices for the $\gamma$ matrices, but, whatever choice is made, there will be mixing of the energy ($\gamma^0 \partial_0$) and mass terms, and a situation in which some of the momentum terms ($\boldsymbol{\gamma}.\boldsymbol{\nabla}$) have real matrix coefficients and others imaginary ones.

Essentially, there are four obvious problems with the matrices:
(1) They cause fragmentation of the equation, mixing up energy, momentum and mass terms.
(2) They take up too much logical space, requiring 16 pieces of information for one operation.
(3) They lack symmetry. There are 5 terms in the equation, but only 4 have a $\gamma$ matrix. Yet there is a fifth matrix ($\gamma^5$) in the algebra.
(4) Even more significantly, as we have seen, the momentum operators are made asymmetric; giving one of the momentum operators an imaginary representation means that our phase space has two spacelike and two timelike components, rather than the 3+1 structure that we believe represents physical reality. Ultimately, this leads to singularities and redundancy on a massive scale, which cannot be fully realised until we have found an alternative formalism which removes them.

There is, of course, no need to use matrices at all, other than historical precedent; and, if we use simpler algebraic operators, we can *defragment* the equation, that is, separate energy, momentum and mass terms from each other, each in its own 'bin', in the same way as we defragment the real and imaginary parts in physical equations using ordinary complex numbers. We will also see that using such operators also solves the logical space problem by reducing the 16 operators of each 4 × 4 matrix to a single term. The only requirement is to find a system of five operators in which
$$(\gamma^0)^2 = (\gamma^5)^2 = 1 \quad (\gamma^1)^2 = (\gamma^2)^2 = (\gamma^3)^2 = -1$$
and *all* terms anticommute with each other, so that $\gamma^0\gamma^1 = -\gamma^1\gamma^0$, etc

The most elegant way of achieving this result might be to use geometrical or Clifford algebra. However, a more *physically* expressive option is to use a combination of quaternions and multivariate vectors, as outlined respectively in (2) and (3), especially as we know that the work of Hestenes and his school has demonstrated that quantum mechanics requires its vector terms to be multivariate, or 'quaternionic', and that this requirement is the origin of the otherwise mysterious property of spin.[3-4]

Preserving the separate identities of the quaternion and vector components, we generate a 32-part algebra, exactly isomorphic to that of the $\gamma$ matrices, with 2 complex scalars, 6 complex vectors, 6 complex quaternions, and 18 complex vector quaternions; and we can easily relate the two algebras by making mappings of the form:

$$\begin{aligned} \gamma^0 &= -i\mathbf{i} \\ \gamma^1 &= \mathbf{i}\mathbf{k} \\ \gamma^2 &= \mathbf{j}\mathbf{k} \\ \gamma^3 &= \mathbf{k}\mathbf{k} \\ \gamma^5 &= i\mathbf{j}, \end{aligned} \qquad (4)$$

or, alternatively,

$$\begin{aligned} \gamma^0 &= i\mathbf{k} \\ \gamma^1 &= i\mathbf{i} \\ \gamma^2 &= \mathbf{j}\mathbf{i} \\ \gamma^3 &= \mathbf{k}\mathbf{i} \\ \gamma^5 &= i\mathbf{j}. \end{aligned} \qquad (5)$$

Both of these mappings generate the full 32-part algebra; and both are relevant to the construction of a defragmented Dirac equation. If, for example, we substitute (4) into the component form of the Dirac equation,

$$\left( \gamma^0 \frac{\partial}{\partial t} + \gamma^1 \frac{\partial}{\partial x} + \gamma^2 \frac{\partial}{\partial y} + \gamma^3 \frac{\partial}{\partial z} + im \right) \psi = 0, \qquad (6)$$

we obtain:

$$\left( -i\mathbf{i} \frac{\partial}{\partial t} + \mathbf{k}\mathbf{i} \frac{\partial}{\partial x} + \mathbf{k}\mathbf{j} \frac{\partial}{\partial y} + \mathbf{k}\mathbf{k} \frac{\partial}{\partial z} + im \right) \psi = 0. \qquad (7)$$

A key move now is to multiply the equation from the left by $\mathbf{j}$, altering the representation to (5), and obtaining

$$\left( i\mathbf{k} \frac{\partial}{\partial t} + i\mathbf{i} \frac{\partial}{\partial x} + i\mathbf{j} \frac{\partial}{\partial y} + i\mathbf{k} \frac{\partial}{\partial z} + ijm \right) \psi = 0. \qquad (8)$$

This apparently trivial step has profound consequences. The equation is now fully symmetrical, and the quaternion operators provide the 3 separated 'bins' we require:

| $\mathbf{k}$ | $\mathbf{i}$ | $\mathbf{j}$ |
|---|---|---|
| energy | momentum | mass |

Although they are mathematically compatible, the new equation (8) takes us well beyond the conventional one. We can see this by applying the plane wave solution for a free-particle

$$\psi = A\, e^{-i(Et - \mathbf{p.r})} \ .$$

where $A$ is the amplitude and $e^{-i(Et - \mathbf{p.r})}$ the phase. We then find that:

$$(kE + i\mathbf{ii}p_x + i\mathbf{ij}p_y + i\mathbf{ik}p_x + ij\, m)\, A\, e^{-i(Et - \mathbf{p.r})} = 0 \ ,$$

which may be more conveniently written in the form,

$$(kE + i\mathbf{i}\,\mathbf{p} + ij\, m)\, A\, e^{-i(Et - \mathbf{p.r})} = 0 \ , \qquad (9)$$

where $\mathbf{p}$ is a multivariate vector.

Here, $(kE + i\mathbf{i}\,\mathbf{p} + ij\, m)$ has a special property. It is a *nilpotent*, or square root of zero, because

$$(kE + i\mathbf{i}\,\mathbf{p} + ij\, m)(kE + i\mathbf{i}\,\mathbf{p} + ij\, m) = -E^2 + p^2 + m^2 = 0 \ . \qquad (10)$$

The only way in which this can be nontrivially accomplished is if $A$ itself is the *same nilpotent*; and, because equation (8) was obtained from equation (7) only by multiplying from the left, the amplitude of the Dirac wavefunction, even within the conventional form of the Dirac equation, must be equally nilpotent, or nilpotent, subject to multiplication by any factor from the right. The derivation, of course, relies on the fact, that, for a multivariate $\mathbf{p}$, the product $\mathbf{pp}$ has a meaning identical to the product of the scalar magnitudes $pp = p^2$. It is also identical to the product of the helicities $(\boldsymbol{\sigma}.\mathbf{p})(\boldsymbol{\sigma}.\mathbf{p})$, which in the case of a fermionic state with positive energy is equal to $(-p)(-p)$, indicating that the multivariate vectors (as equivalent to Pauli matrices) automatically incorporate the concept of spin. (While using a multivariate vector $\mathbf{p}$ means that we obtain the spin directly, without needing $\boldsymbol{\sigma}.\mathbf{p}$, in cases where, for mathematical convenience, we reduce it to an ordinary vector, the spin must be explicitly introduced.)

**4 The Dirac 4-spinor**

Of course, $\psi$ is not really a single term, but a 4-component spinor, which accommodates fermion and antifermion states, as well as spin up and spin down. However, identification of these is now easy:

| | |
|---|---|
| fermion / antifermion | $\pm E$ |
| spin up / down | $\pm \mathbf{p}$ |

We now need a 4-spinor with 4 amplitudes and 4 phases, with all 4 variations of $\pm E$ and $\pm \mathbf{p}$ applied to $(kE + i\mathbf{i}\,\mathbf{p} + i\mathbf{j}\,m)\,e^{-i(Et - \mathbf{p} \cdot \mathbf{r})}$.

The Dirac 4-spinor is now a column vector with 4 components:

$$\psi_1 = (kE + i\mathbf{i}\,\mathbf{p} + i\mathbf{j}\,m)\,e^{-i(Et - \mathbf{p} \cdot \mathbf{r})}$$
$$\psi_2 = (kE - i\mathbf{i}\,\mathbf{p} + i\mathbf{j}\,m)\,e^{-i(Et + \mathbf{p} \cdot \mathbf{r})}$$
$$\psi_3 = (-kE + i\mathbf{i}\,\mathbf{p} + i\mathbf{j}\,m)\,e^{i(Et - \mathbf{p} \cdot \mathbf{r})}$$
$$\psi_4 = (-kE - i\mathbf{i}\,\mathbf{p} + i\mathbf{j}\,m)\,e^{i(Et + \mathbf{p} \cdot \mathbf{r})}\ .$$

each of which is operated on by $(i k \partial/\partial t + i \nabla + i\mathbf{j}\,m)$.

However, there is one more trick we can play, and it is essential to obtaining a fully defragmented Dirac equation. Since our differential operator has been reduced to a single term from the 16 in the original matrix, we now have the logical space to turn it into a 4-spinor, like the wavefunction, and reduce to a *single phase*. That is, we transfer the variation in the signs of $E$ and $\mathbf{p}$ from the exponential to the differential operator. So, we now have:

$$\left(ik\frac{\partial}{\partial t} + i\nabla + ijm\right)(kE + ii\mathbf{p} + ijm)e^{-i(Et - \mathbf{p} \cdot \mathbf{r})} = 0$$

$$\left(ik\frac{\partial}{\partial t} - i\nabla + ijm\right)(kE - ii\mathbf{p} + ijm)e^{-i(Et - \mathbf{p} \cdot \mathbf{r})} = 0$$

$$\left(-ik\frac{\partial}{\partial t} + i\nabla + ijm\right)(-kE + ii\mathbf{p} + ijm)e^{-i(Et - \mathbf{p} \cdot \mathbf{r})} = 0$$

$$\left(-ik\frac{\partial}{\partial t} - i\nabla + ijm\right)(-kE - ii\mathbf{p} + ijm)e^{-i(Et - \mathbf{p} \cdot \mathbf{r})} = 0$$

The Dirac 4-spinor equation for a free particle can now be represented by a row vector of 4 differential operators acting on a column vector of 4 eigenstates. Using a compactified notation:

$$\left(\pm ik\frac{\partial}{\partial t} \pm i\nabla + ijm\right)(\pm kE \pm ii\mathbf{p} + ijm)e^{-i(Et - \mathbf{p} \cdot \mathbf{r})} = 0 \qquad (11)$$

*row*                *column*

This formalism automatically includes the Feynman representation of antistates with negative energy going backwards in time. It ensures that negative energy only occurs within negative time, and that all states defined in a time compatible with thermodynamics are positive energy states. It follows also that positive energy states require positive mass states, if we define mass as the 'proper energy', or energy within the fermion's inertial frame, in the same way as the (positive) proper time $\tau$ in the conjugate nilpotent expression $(\pm kt \pm i\mathbf{i}\,\mathbf{r} + i\mathbf{j}\,\tau)$, is defined as the time within the fermion's inertial frame.

If we had written (10) in spinor form, we could, of course, have derived (11) from it directly, by converting the $E$ and $\mathbf{p}$ terms into quantum operators, and we could use it in the same way to derive the Klein-Gordon equation, which, in this formalism, becomes merely a branch of Dirac, with an additional conversion of the amplitude:

$$\left(\pm i\mathbf{k}\frac{\partial}{\partial t} \pm \mathbf{i}\nabla + \mathbf{ij}m\right)\left(\pm i\mathbf{k}\frac{\partial}{\partial t} \pm \mathbf{i}\nabla + \mathbf{ij}m\right)e^{-i(Et-\mathbf{p}\cdot\mathbf{r})} = 0$$

However, while this equation can be applied to a wavefunction of any kind, for example the scalar wavefunction of a boson, because the nilpotency is provided by the double differential operator, (11) can only be applied to a state which is a nilpotent.

An important qualification must be made, however, to the formalism as presented in (11). This is derived directly from the Dirac equation as normally presented. But, from a *physical* point of view, it would be more correct to write the nilpotent operator in the form

$$\left(\pm i\mathbf{k}E \pm \mathbf{i}\mathbf{p} + \mathbf{j}m\right)$$

in which the energy operator becomes a pseudoscalar, and the momentum and mass operators remain real. Though the form ($\pm \mathbf{k}E \pm \mathbf{ii}\,\mathbf{p} + \mathbf{ij}\,m$) will be adopted as a convention for most of this account, it will always be assumed that, *physically*, energy, and not mass, is pseudoscalar.

Using either convention, it may be said, the equation in the defragmented nilpotent form has also reached its most perfect pitch of simplicity and symmetry. Operator and amplitude are essentially identical, since

$$\left(\pm \mathbf{k}E \pm \mathbf{ii}\mathbf{p} + \mathbf{ij}m\right) \quad \text{is just} \quad \left(\pm i\mathbf{k}\frac{\partial}{\partial t} \pm \mathbf{i}\nabla + \mathbf{ij}m\right)$$

in operator form. In fact, we can even do away with the equation altogether! All we need is to specify the operator meanings for the terms separated by the three quaternion 'bins':

$$(\pm \mathbf{k}E \pm \mathbf{ii}\,\mathbf{p} + \mathbf{ij}\,m)$$

That is, $E$ and $\mathbf{p}$ don't need to be the differential operators for a free state. They can be covariant derivatives or incorporate field terms of any kind. If $E$ and $\mathbf{p}$ contain field terms, then the phase will no longer be a pure exponential $e^{-i(Et-\mathbf{p}\cdot\mathbf{r})}$. It will be whatever function is needed to make the amplitude nilpotent. Both the amplitude and phase terms will be uniquely defined once the differential operator has been specified. And because there is only one phase term, analytical solutions will be easier to find.

Again, although the operator is in principle a 4-component spinor, the terms are not independent, and all the information is contained in the lead term. The remaining terms then simply offer a fixed pattern of sign variations for the energy and momentum components; and the total effect of this variation is merely to ensure that fermion amplitudes have only two possible products with each other after normalization: 0 if they are identical and 1 if they are not (almost like the reverse of a delta function). Essentially, then, although we will continue to use four components for the maximum clarity, the fermionic state, and all calculation related to it, can be reduced to a single-line operator, composed of an energy term, a momentum term and a mass term. Symbolically, we have $ikE + ii\mathbf{p} + ijm$ or $ikE + i\mathbf{p} + jm$, where $E$ and $\mathbf{p}$ are either operators or eigenvalues.

**5 Bosonic states**

Even more important than obtaining particular solutions is the examination of the Dirac state (or operator) itself. In principle, it should contain all the information about fermions (and bosons) that can exist. In this case, it should be a unique key to explaining fundamental physical facts. The nilpotent Dirac state is, of course, a quantum operator. So it is useful to show that it can do conventional quantum mechanics. Here, we can define a probability density by multiplying the amplitude by its complex quaternion conjugate $(\pm kE \pm ii\mathbf{p} + ijm)(\mp kE \pm ii\mathbf{p} + ijm)$, which produces the positive definite value $8E^2$, and normalising to unity, though, because of the nature of nilpotent mathematics, we will seldom need to make the normalisation explicit. Using this definition, the 'reciprocal' of ($\pm kE \pm ii\,\mathbf{p} + ij\,m$) that occurs, for example, in the propagator, can be identified as $(\mp kE \pm ii\mathbf{p} + ijm)$.

The nilpotent state, however, is superior to the conventional quantum state – it is automatically second quantized, with in-built supersymmetry. Amplitude and phase are uniquely determined by the same operator and each is quantized in the same way. Formal second quantization is unnecessary. In this formulation, the differential operator and the eigenvalue part of the wavefunction are essentially identical, and are quantized in identical ways. The reason for writing one as a differential operator and one as an eigenvalue term is that we then have a simultaneous representation of the nonconserved and the conserved parts of the equation; the Dirac equation for a free particle can then be seen as a mathematical expression of the absolute conservation of $E$, $\mathbf{p}$ and $m$ and the absolute *nonconservation*, or variability, of $\mathbf{r}$ and $t$. In this version of the equation, also, as we have seen, the four solutions exist at the same time and on the same footing – they differ only in signs of $E$ and $\mathbf{p}$. Quantum field integrals acting on vacuum produce the nilpotent state vector.[5]

Essentially, then, we have the basic requirements for automatic second quantization and a quantum field theory, the nilpotent expressions displaying the characteristics of full quantum field operators rather than wavefunctions in the more restricted sense. This means that creation / annihilation operators are easily identified. The fermion state ($\pm kE \pm ii\mathbf{p} + ijm$) incorporates 4 creation (or annihilation) operators:

| | |
|---|---|
| Fermion creation spin up | ($kE + ii\mathbf{p} + ijm$) |
| Fermion creation spin down | ($kE - ii\mathbf{p} + ijm$) |
| Antifermion creation spin down | ($- kE + ii\mathbf{p} + ijm$) |
| Antifermion creation spin up | ($- kE - ii\mathbf{p} + ijm$) |

| Antifermion annihilation spin down | $(kE + i\mathbf{i}\mathbf{p} + ijm)$ |
| Antifermion annihilation spin up | $(kE - i\mathbf{i}\mathbf{p} + ijm)$ |
| Fermion annihilation spin up | $(-kE + i\mathbf{i}\mathbf{p} + ijm)$ |
| Fermion annihilation spin down | $(-kE - i\mathbf{i}\mathbf{p} + ijm)$ |

We can also immediately recognize the state vectors for

| Fermion | $(\pm kE \pm i\mathbf{i}\mathbf{p} + ijm)$ |
| Fermion with reversed spin | $(\pm kE \mp i\mathbf{i}\mathbf{p} + ijm)$ |
| Antifermion | $(\mp kE \pm i\mathbf{i}\mathbf{p} + ijm)$ |
| Antifermion with reversed spin | $(\mp kE \mp i\mathbf{i}\mathbf{p} + ijm)$ |

So we can proceed to construct state vectors for spin 1 bosons, spin 0 bosons and the kind of combinations that produce Bose-Einstein condensates, as, respectively, scalar products of fermion / antifermion with the same helicity; fermion / antifermion with opposite helicity; and fermion / fermion with opposite helicity. All possible interaction vertices between one fermion / antifermion state and another are scalar quantities, as all nonzero products of nilpotents are scalar. Real bosonic *states* are formed when the energy and momentum values at a vertex are equalized between the two states, and states can be specified as follows:

| Spin 1 boson | $(\pm kE \pm i\mathbf{i}\mathbf{p} + ijm)\ (\mp kE \pm i\mathbf{i}\mathbf{p} + ijm)$ |
| Spin 0 boson | $(\pm kE \pm i\mathbf{i}\mathbf{p} + ijm)\ (\mp kE \mp i\mathbf{i}\mathbf{p} + ijm)$ |
| Bose-Einstein condensate | $(\pm kE \pm i\mathbf{i}\mathbf{p} + ijm)\ (\pm kE \mp i\mathbf{i}\mathbf{p} + ijm)$ |

We notice here that the boson is structured as a unified state, with *E*, **p** and *m* values common to the fermionic and antifermionic parts. We can, in fact, postulate that the signature of a completely interacting dynamical theory of composite particles is that the *E*, **p** and *m* values have meaning only in the context of the *entire state*. This will become especially significant with baryons.

Expanding upon the above, if we represent a fermion by the row vector:

$(kE + i\mathbf{i}\ \mathbf{p} + ij\ m)$
$(kE - i\mathbf{i}\ \mathbf{p} + ij\ m)$
$(-kE + i\mathbf{i}\ \mathbf{p} + ij\ m)$
$(-kE - i\mathbf{i}\ \mathbf{p} + ij\ m)$

and an antifermion of the same helicity by the column vector:

$(-kE + i\mathbf{i}\ \mathbf{p} + ij\ m)$
$(-kE - i\mathbf{i}\ \mathbf{p} + ij\ m)$
$(kE + i\mathbf{i}\ \mathbf{p} + ij\ m)$
$(kE - i\mathbf{i}\ \mathbf{p} + ij\ m)\,,$

a spin 1 boson can be represented by

$$(kE + i\boldsymbol{i}\,\mathbf{p} + i\boldsymbol{j}\,m) \quad (-kE + i\boldsymbol{i}\,\mathbf{p} + i\boldsymbol{j}\,m)$$
$$(kE - i\boldsymbol{i}\,\mathbf{p} + i\boldsymbol{j}\,m) \quad (-kE - i\boldsymbol{i}\,\mathbf{p} + i\boldsymbol{j}\,m)$$
$$(-kE + i\boldsymbol{i}\,\mathbf{p} + i\boldsymbol{j}\,m) \quad (kE + i\boldsymbol{i}\,\mathbf{p} + i\boldsymbol{j}\,m)$$
$$(-kE - i\boldsymbol{i}\,\mathbf{p} + i\boldsymbol{j}\,m) \quad (kE - i\boldsymbol{i}\,\mathbf{p} + i\boldsymbol{j}\,m),$$

which sums to the scalar value $4(E^2 + p^2 + m^2) = 8E^2$ before normalisation. Even if the boson is massless, the product will be the same, since then $4(E^2 + p^2) = 8E^2$. Massless spin 1 bosons are, of course, the key mediators for the strong and electric interactions (gluons, photons).

In the same way, if we represent a spin 0 boson by

$$(kE + i\boldsymbol{i}\,\mathbf{p} + i\boldsymbol{j}\,m) \quad (-kE - i\boldsymbol{i}\,\mathbf{p} + i\boldsymbol{j}\,m)$$
$$(kE - i\boldsymbol{i}\,\mathbf{p} + i\boldsymbol{j}\,m) \quad (-kE + i\boldsymbol{i}\,\mathbf{p} + i\boldsymbol{j}\,m)$$
$$(-kE + i\boldsymbol{i}\,\mathbf{p} + i\boldsymbol{j}\,m) \quad (kE - i\boldsymbol{i}\,\mathbf{p} + i\boldsymbol{j}\,m)$$
$$(-kE - i\boldsymbol{i}\,\mathbf{p} + i\boldsymbol{j}\,m) \quad (kE + i\boldsymbol{i}\,\mathbf{p} + i\boldsymbol{j}\,m),$$

we will obtain the scalar value $4(E^2 - p^2 + m^2) = 8m^2$ before normalisation. However, this time there is a significant change. If the boson is massless, the product will be zero, since then $4(E^2 - p^2) = 0$. Here, we can also use the concept of Pauli exclusion, since any product of the form

$$(\pm kE \pm i\boldsymbol{i}\mathbf{p} + i\boldsymbol{j}m) \quad (\pm kE \pm i\boldsymbol{i}\mathbf{p} + i\boldsymbol{j}m)$$

between identical fermions will always be zero for a nilpotent state vector. The same will apply to $(\pm kE \pm i\boldsymbol{i}\mathbf{p})\ (\mp kE \mp i\boldsymbol{i}\mathbf{p})$ for a massless spin 0 boson. In the nilpotent formalism, massless spin 0 bosons (for example, Goldstone bosons) are mathematically impossible.

The third option we have considered, the 'Bose-Einstein condensate', also multiplies to a nonzero scalar ($-8p^2$ before normalisation).

$$(kE + i\boldsymbol{i}\,\mathbf{p} + i\boldsymbol{j}\,m) \quad (kE - i\boldsymbol{i}\,\mathbf{p} + i\boldsymbol{j}\,m)$$
$$(kE - i\boldsymbol{i}\,\mathbf{p} + i\boldsymbol{j}\,m) \quad (kE + i\boldsymbol{i}\,\mathbf{p} + i\boldsymbol{j}\,m)$$
$$(-kE + i\boldsymbol{i}\,\mathbf{p} + i\boldsymbol{j}\,m) \quad (-kE - i\boldsymbol{i}\,\mathbf{p} + i\boldsymbol{j}\,m)$$
$$(-kE - i\boldsymbol{i}\,\mathbf{p} + i\boldsymbol{j}\,m) \quad (-kE + i\boldsymbol{i}\,\mathbf{p} + i\boldsymbol{j}\,m).$$

We can identify this, at least in an idealised form, as being the spin 0 state we would expect from: the unit bosonic state in a Bose-Einstein condensate or an even-even nucleus; the Cooper pairs in a superconductor (e.g. $He^4$); the combination of electron and magnetic flux line in the quantum Hall phenomenon; the single-valued wavefunctions produced in other applications of the Berry phase, such as the Aharonov-Bohm effect and the Jahn-Teller effect. Many of these are spin 0 combinations, but there is one circumstance in which we would expect something different. This is if the two fermionic components are physically separated in such a way that their momentum components can be oppositely aligned, but with the same direction of spin. The obvious example is $He^3$, which is a superconductor with spin 1.

As mentioned previously, the product of two identical fermions vanishes
$$(kE + i\mathbf{i}\,\mathbf{p} + i\mathbf{j}\,m)\quad (kE + i\mathbf{i}\,\mathbf{p} + i\mathbf{j}\,m) = 0$$
$$(kE - i\mathbf{i}\,\mathbf{p} + i\mathbf{j}\,m)\quad (kE - i\mathbf{i}\,\mathbf{p} + i\mathbf{j}\,m) = 0$$
$$(-kE + i\mathbf{i}\,\mathbf{p} + i\mathbf{j}\,m)\quad (kE + i\mathbf{i}\,\mathbf{p} + i\mathbf{j}\,m) = 0$$
$$(-kE - i\mathbf{i}\,\mathbf{p} + i\mathbf{j}\,m)\quad (kE - i\mathbf{i}\,\mathbf{p} + i\mathbf{j}\,m) = 0$$

The Pauli exclusion principle in this form demands nonlocality. It also means that Pauli exclusion can be established for each individual fermion, in calculations on many fermion system, without using the entire Slater determinant.

It will be evident, in addition, that nilpotent operators are naturally supersymmetric, with supersymmetry operators:

Boson to fermion: $\quad Q = (\pm kE \pm i\mathbf{i}\mathbf{p} + i\mathbf{j}m)$

Fermion to boson: $\quad Q^\dagger = (\mp kE \pm i\mathbf{i}\mathbf{p} + i\mathbf{j}m)$

So we can use $Q$ and $Q^\dagger$ to convert bosons to fermions / antifermions and vice versa. The supersymmetry is exact. Such exact supersymmetry suggests that particles are their own supersymmetric partners. This will become apparent when we study vacuum.

**6 CPT Symmetry**

There are three fundamental symmetry operations in particle physics:
- $P$      Parity               reverses signs of space coordinates
- $T$      Time reversal     reverses sign of time coordinate
- $C$      Charge conjugation   exchanges particle and antiparticle

The laws of physics are not preserved under these transformations taken separately, but are preserved under all three operations taken together (*CPT*). *CPT* symmetry is another mathematical consequences of a nilpotent representation. We can represent the component *P*, *T* and *C* operations on a nilpotent wavefunction by using a different operator to represent each type of transformation:

$P$: $\quad \mathbf{i}(\pm kE \pm i\mathbf{i}\mathbf{p} + i\mathbf{j}m)\mathbf{i} = (\pm kE \mp i\mathbf{i}\mathbf{p} + i\mathbf{j}m)$

$T$: $\quad \mathbf{k}(\pm kE \pm i\mathbf{i}\mathbf{p} + i\mathbf{j}m)\mathbf{k} = (\mp kE \pm i\mathbf{i}\mathbf{p} + i\mathbf{j}m)$

$C$: $\quad -\mathbf{j}(\pm kE \pm i\mathbf{i}\mathbf{p} + i\mathbf{j}m)\mathbf{j} = (\mp kE \mp i\mathbf{i}\mathbf{p} + i\mathbf{j}m)$

The last may also be written:

$C$: $\quad i\mathbf{j}(\pm kE \pm i\mathbf{i}\mathbf{p} + i\mathbf{j}m)i\mathbf{j} = (\mp kE \mp i\mathbf{i}\mathbf{p} + i\mathbf{j}m)$

From this we may see that:

$CP = T$: $\quad -\mathbf{j}(\mathbf{i}(\pm kE \pm i\mathbf{i}\mathbf{p} + i\mathbf{j}m)\mathbf{i})\mathbf{j} = \mathbf{k}(\pm kE \pm i\mathbf{i}\mathbf{p} + i\mathbf{j}m)\mathbf{k} = (\mp kE \pm i\mathbf{i}\mathbf{p} + i\mathbf{j}m)$

$PT = C$: $\quad \mathbf{i}(\mathbf{k}(\pm kE \pm i\mathbf{i}\mathbf{p} + i\mathbf{j}m)\mathbf{k})\mathbf{i} = -\mathbf{j}(\pm kE \pm i\mathbf{i}\mathbf{p} + i\mathbf{j}m)\mathbf{j} = (\mp kE \mp i\mathbf{i}\mathbf{p} + i\mathbf{j}m)$

$TC = P$: $\quad \mathbf{k}(-\mathbf{j}(\pm kE \pm i\mathbf{i}\mathbf{p} + i\mathbf{j}m)\mathbf{j})\mathbf{k} = \mathbf{i}(\pm kE \pm i\mathbf{i}\mathbf{p} + i\mathbf{j}m)\mathbf{i} = (\pm kE \mp i\mathbf{i}\mathbf{p} + i\mathbf{j}m)$

and that $TCP \equiv$ identity, because:

$$k(-j(i(\pm kE \pm i i\mathbf{p} + ijm)i)j)k = -kji(\pm kE \pm ii\mathbf{p} + ijm)ijk = (\pm kE \pm ii\mathbf{p} + ijm)$$

**7 Spin**

It is important, at this stage, to establish some fundamental results obtained by conventional means before exploring the special characteristics of the new formalism. Particularly important is the derivation of fermion spin. In the conventional treatment of spin, we define a vector $\hat{\sigma}$, with components

$$\hat{\sigma}_l = i\gamma_0\gamma_5\gamma_l, \text{ with } l = 1, 2, 3$$

Now, in the nilpotent formalism

$$\gamma_0 = i\mathbf{k} \; ; \; \gamma_1 = i\mathbf{i} \; ; \; \gamma_2 = j\mathbf{i} \; ; \; \gamma_3 = \mathbf{k}i \; ; \; \gamma_5 = ij.$$

So,

$$\hat{\sigma}_1 = -\mathbf{i} \; ; \; \hat{\sigma}_2 = -\mathbf{j} \; ; \; \hat{\sigma}_3 = -\mathbf{k}$$

or, in more convenient notation,

$$\hat{\sigma} = -\mathbf{1},$$

and

$$\gamma = i\mathbf{1}.$$

Suppose, now, that $\mathbf{L}$ is the orbital angular momentum of a fermion, $\mathbf{r} \times \mathbf{p}$. Then, in the standard formalism,

$$[\mathbf{L}, \mathcal{H}] = [\mathbf{r} \times \mathbf{p}, i\gamma_0\gamma.\mathbf{p} + \gamma_0 m] = [\mathbf{r} \times \mathbf{p}, i\gamma_0\gamma.\mathbf{p}].$$

Taking out common factors,

$$[\mathbf{L}, \mathcal{H}] = i\gamma_0 [\mathbf{r}, \gamma.\mathbf{p}] \times \mathbf{p}$$

In the nilpotent algebra, this is equivalent to

$$[\mathbf{L}, \mathcal{H}] = -\mathbf{k}i [\mathbf{r}, \mathbf{1}.\mathbf{p}] \times \mathbf{p} = -j [\mathbf{r}, \mathbf{1}.\mathbf{p}] \times \mathbf{p}.$$

Now,

$$[\mathbf{r},\mathbf{1}.\mathbf{p}]\psi = -i\mathbf{i}\left(x\frac{\partial\psi}{\partial x} - \frac{\partial(x\psi)}{\partial x}\right) - i\mathbf{j}\left(y\frac{\partial\psi}{\partial y} - \frac{\partial(y\psi)}{\partial y}\right) - i\mathbf{k}\left(z\frac{\partial\psi}{\partial z} - \frac{\partial(z\psi)}{\partial z}\right) = i\mathbf{1}\psi.$$

Hence,

$$[\mathbf{L}, \mathcal{H}] = -ij\,\mathbf{1} \times \mathbf{p}. \tag{12}$$

So, $\mathbf{L}$ is not a constant of the motion, without the additional 'spin' term $ij\,\mathbf{1} \times \mathbf{p}$.

Suppose we now consider the expression

$$[\hat{\sigma}, \mathcal{H}] = [\hat{\sigma}, i\gamma_0\gamma.\mathbf{p} + \gamma_0 m]$$
$$[-\mathbf{1}, \mathcal{H}] = [-\mathbf{1}, -j\,(\mathbf{i}p_1 + \mathbf{j}p_2 + \mathbf{k}p_3) + i\mathbf{k}m]$$
$$= [-\mathbf{1}, -j\,(\mathbf{i}p_1 + \mathbf{j}p_2 + \mathbf{k}p_3)]$$

since *ikm* and −**1** commute. Multiplying this out, we obtain
$$[-\mathbf{1}, \mathcal{H}] = 2\boldsymbol{j}\,(\mathbf{ij}p_2 + \mathbf{ik}p_3 + \mathbf{ji}p_1 + \mathbf{jk}p_3 + \mathbf{ki}p_1 + \mathbf{kj}p_2)$$
$$= 2\boldsymbol{ij}\,(\mathbf{k}(p_2 - p_1) + \mathbf{j}(p_1 - p_3) + \mathbf{i}(p_3 - p_2))$$
$$= 2\boldsymbol{ij}\,\mathbf{1} \times \mathbf{p}\,. \tag{13}$$
Combining (12) and (13), we obtain
$$[\mathbf{L} - \mathbf{1}\,/\,2,\, \mathcal{H}] = 0$$
which is equivalent to the conventional
$$[\mathbf{L} + \hat{\boldsymbol{\sigma}}\,/\,2,\, \mathcal{H}] = 0\,.$$
Hence, $(\mathbf{L} - \mathbf{1}\,/\,2) = (\mathbf{L} + \hat{\boldsymbol{\sigma}}\,/\,2)$ is a constant of the motion.

The nilpotent formulation provides a ready explanation for the spin-statistics connection. The ½-value of fermion spin, which comes from the noncommutativity of the components of **p** in the usual formal derivations, may be taken as a consequence of the fact that the fermion wavefunction is a square root or a single noncommutative nilpotent, while the 0 or 1 value of boson spin is a consequence of the boson wavefunction being a commutative (scalar) product of two square roots or nilpotents.

## 8 Helicity

The term
$$\hat{\boldsymbol{\sigma}}\cdot\mathbf{p} = -\mathbf{1}\cdot\mathbf{p} = -p_1 - p_2 - p_3 = -p$$
is defined as helicity, and, since it has no vector or quaternion terms, and has only terms of the form $\partial/\partial x$, $\partial/\partial y$, and $\partial/\partial z$ in common with
$$i\gamma_0\boldsymbol{\gamma}\cdot\mathbf{p} = -\boldsymbol{j}\,(\mathbf{i}p_1 + \mathbf{j}p_2 + \mathbf{k}p_3)$$
and also clearly commutes with $\gamma_0 m = i\boldsymbol{k}m$, then
$$[\hat{\boldsymbol{\sigma}}\cdot\mathbf{p},\, \mathcal{H}] = [-\mathbf{1}\cdot\mathbf{p},\, \mathcal{H}] = 0$$
and the helicity is a constant of the motion.

For a hypothetical particle with zero mass, the term $\boldsymbol{k}E + i\boldsymbol{i}p + \boldsymbol{ij}\,m$ reduces to $\boldsymbol{k}E + i\boldsymbol{i}p$, where $p$ actually represents $\hat{\boldsymbol{\sigma}}\cdot\mathbf{p} = -\mathbf{1}\cdot\mathbf{p}$. Numerically, $E$ also becomes equal to $\pm p$. For positive energy states,
$$E = \hat{\boldsymbol{\sigma}}\cdot\mathbf{p} = -\mathbf{1}\cdot\mathbf{p}\,.$$
So the spin is aligned antiparallel to the momentum (has left-handed helicity). Then,
$$\boldsymbol{ij}\,(\boldsymbol{k}E + i\boldsymbol{i}p) = \boldsymbol{ij}\,(\boldsymbol{k} - i\boldsymbol{i}\,)\,E = (i\boldsymbol{i} - \boldsymbol{k})\,E$$
and the spinor wavefunction follows the rule:
$$\boldsymbol{ij}\,u_L = -u_L\,. \tag{14}$$
For negative energy states,
$$E = -\hat{\boldsymbol{\sigma}}\cdot\mathbf{p}\,,$$
In this case, the spin is aligned parallel to the momentum (has right-handed helicity). Then,
$$\boldsymbol{ij}\,(\boldsymbol{k}E + i\boldsymbol{i}p) = \boldsymbol{ij}\,(\boldsymbol{k} + i\boldsymbol{i}\,)\,E = (i\boldsymbol{i} + \boldsymbol{k})\,E$$
and the spinor wavefunction follows the rule:

$$ij\, u_R = u_R .\tag{15}$$

Equations (14) and (15) allow $ij = \gamma_5$ to be used, in conventional theory, as a projection operator. From these equations, we may derive the relations

$$\left(\frac{1-ij}{2}\right)u_L = \left(\frac{1-\gamma_5}{2}\right)u_L = u_L$$

and

$$\left(\frac{1-ij}{2}\right)u_R = \left(\frac{1-\gamma_5}{2}\right)u_R = 0 .$$

These are the equations which, in conventional theory, produce two sharply-defined helicity states ($\sigma.\mathbf{p} / 2p = \frac{1}{2}$ and $-\frac{1}{2}$), of which the right-handed state ($p$) is suppressed in the case of assumed massless fermions of positive energy ($E$) and the left-handed state ($-p$) in the case of massless fermions of 'negative energy' ($-E$), or antifermions, in which $E = -p$. Helicity is a pseudoscalar and so changes sign under parity transformations; this means that parity must be violated in interactions such as those involving massless fermions because the two helicity states do not then make equal contributions to the interaction. Parity violation, in this case, is made inevitable by the suppression of mass, and the fixing of the $E / p$ ratio.

A more fundamental way of looking at the problem is to return to the question of spin 1 and spin 0 bosons. Nilpotent structure ensures that spin 1 bosons can be massless, while spin 0 bosons cannot. Only a massless particle can be exclusively one-handed. The spin 1 boson has components that may be exclusively one-handed because if the spins of fermion and antifermion are aligned (in the **p** component), their helicities will be opposite because of their opposite signs of $E$. A spin 0 boson, however, has fermion and antifermion components that are in opposite spin (or **p**) alignment, so with states of the *same* helicity because of their opposite signs of $E$. The fact that nilpotency determines that spin 0 bosons must have nonzero mass ensures that this is equivalent to having fermion and antifermion states of the same helicity, while masslessness requires states of opposite helicity. As we have seen, the anticommutativity of the operations involved in defining [**L**, $\mathcal{H}$] ultimately ensure that the helicity term is antiparallel (or left-handed) for positive energy states and parallel (or right-handed) for negative energy states.

**9 Baryons**

We have already postulated an entangled system of two nilpotent states (fermion and antifermion) to describe bosons. Can we extend this idea to three nilpotent states to describe baryons? Conventionally, we consider a baryon to be made up of three fermionic components, to which we assign colour to overcome Pauli exclusion. Can we relate this concept of colour to the fundamental structure of nilpotents? Can we have a 3-component state vector? Obviously a combination involving identical fermions will be impossible, because

$$(kE + i\mathbf{i}\,\mathbf{p} + i\mathbf{j}\,m)\,(kE + i\mathbf{i}\,\mathbf{p} + i\mathbf{j}\,m)\,(kE + i\mathbf{i}\,\mathbf{p} + i\mathbf{j}\,m) = 0$$

But

$$(kE + i\mathbf{i}\,\mathbf{p} + i\mathbf{j}\,m)\,(kE + i\mathbf{j}\,m)\,(kE + i\mathbf{j}\,m) \rightarrow (kE + i\mathbf{i}\,\mathbf{p} + i\mathbf{j}\,m)$$
$$(kE + i\mathbf{j}\,m)\,(kE + i\mathbf{i}\,\mathbf{p} + i\mathbf{j}\,m)\,(kE + i\mathbf{j}\,m) \rightarrow (kE - i\mathbf{i}\,\mathbf{p} + i\mathbf{j}\,m)$$
$$(kE + i\mathbf{j}\,m)\,(kE + i\mathbf{j}\,m)\,(kE + i\mathbf{i}\,\mathbf{p} + i\mathbf{j}\,m) \rightarrow (kE + i\mathbf{i}\,\mathbf{p} + i\mathbf{j}\,m)$$

after normalization of the scalar factor $-p^2$. Also

$$(kE - i\mathbf{i}\,\mathbf{p} + i\mathbf{j}\,m)\,(kE + i\mathbf{j}\,m)\,(kE + i\mathbf{j}\,m) \rightarrow (kE - i\mathbf{i}\,\mathbf{p} + i\mathbf{j}\,m)$$
$$(kE + i\mathbf{j}\,m)\,(kE - i\mathbf{i}\,\mathbf{p} + i\mathbf{j}\,m)\,(kE + i\mathbf{j}\,m) \rightarrow (kE + i\mathbf{i}\,\mathbf{p} + i\mathbf{j}\,m)$$
$$(kE + i\mathbf{j}\,m)\,(kE + i\mathbf{j}\,m)\,(kE - i\mathbf{i}\,\mathbf{p} + i\mathbf{j}\,m) \rightarrow (kE - i\mathbf{i}\,\mathbf{p} + i\mathbf{j}\,m)$$

(Here, for convenience, we have taken only the first of the four terms of the tensor product.) So it is possible to have a nonzero state vector if we use the *vector* properties of **p** and the arbitrary nature of its sign (+ or –). A state vector of the form, privileging the **p** components:

$$(kE \pm i\mathbf{i}\,\mathbf{i}p_x + i\mathbf{j}\,m)\,(kE \pm i\mathbf{i}\,\mathbf{j}p_y + i\mathbf{j}\,m)\,(kE \pm i\mathbf{i}\,\mathbf{k}p_z + i\mathbf{j}\,m)$$

has six independent allowed phases, i.e. when

$$\mathbf{p} = \pm \mathbf{i}p_x,\ \mathbf{p} = \pm \mathbf{j}p_y,\ \mathbf{p} = \pm \mathbf{k}p_z$$

But these must be *gauge invariant*, i.e. indistinguishable, or all present at once. This requires an exact symmetry with an *SU*(3) group structure, with eight generators, exactly comparable to the conventional symmetry of the coloured quark model, with three symmetric and three antisymmetric phases, and transitions mediated by eight massless spin 1 gluons.

| | |
|---|---|
| $(kE + i\mathbf{i}\,\mathbf{i}p_x + i\mathbf{j}\,m)\,(kE + \ldots + i\mathbf{j}\,m)\,(kE + \ldots + i\mathbf{j}\,m)$ | +RGB |
| $(kE - i\mathbf{i}\,\mathbf{i}p_x + i\mathbf{j}\,m)\,(kE - \ldots + i\mathbf{j}\,m)\,(kE - \ldots + i\mathbf{j}\,m)$ | –RBG |
| $(kE + \ldots + i\mathbf{j}\,m)\,(kE + i\mathbf{i}\,\mathbf{j}p_y + i\mathbf{j}\,m)\,(kE + \ldots + i\mathbf{j}\,m)$ | +BRG |
| $(kE - \ldots + i\mathbf{j}\,m)\,(kE - i\mathbf{i}\,\mathbf{j}p_y + i\mathbf{j}\,m)\,(kE - \ldots + i\mathbf{j}\,m)$ | –GRB |
| $(kE + \ldots + i\mathbf{j}\,m)\,(kE + \ldots + i\mathbf{j}\,m)\,(kE + i\mathbf{i}\,\mathbf{k}p_z + i\mathbf{j}\,m)$ | +GBR |
| $(kE - \ldots + i\mathbf{j}\,m)\,(kE - \ldots + i\mathbf{j}\,m)\,(kE - i\mathbf{i}\,\mathbf{k}p_z + i\mathbf{j}\,m)$ | –BGR   (16) |

$$\psi \sim (RGB - RBG + BRG - GRB + GBR - BGR)$$

If such a structure really does represent a baryon wavefunction, then we can predict that the spin is a property of the baryon wavefunction as a whole, not of component quark wavefunctions. It is, of course, immaterial, with respect to the final result, whether the signs of the absent components of momentum are positive or negative. So, it would be possible to obtain the same patterns with all three signs of $p_x$, $p_y$, $p_z$ in each combination the same or with one different, just as it is possible to have baryons with overall spin ½ or 3/2. It is extremely significant, however, that this whole representation is impossible in a conventional spinor formulation, with terms such as $p_x + ip_y$, or in any representation in which the momentum operators cannot show the full affine nature of the vector concept.

The eight gluon vertices, in this formulation, are constructed from:

$$(\pm k E \mp ii\,\mathbf{i}p_x) (\mp kE \mp ii\,\mathbf{j}p_y) \quad (\pm kE \mp ii\,\mathbf{j}p_y) (\mp kE \mp ii\,\mathbf{i}p_x)$$
$$(\pm kE \mp ii\,\mathbf{j}p_y) (\mp kE \mp ii\,\mathbf{k}p_z) \quad (\pm kE \mp ii\,\mathbf{k}p_z) (\mp kE \mp ii\,\mathbf{j}p_y)$$
$$(\pm kE \mp ii\,\mathbf{i}p_z) (\mp kE \mp ii\,\mathbf{i}p_x) \quad (\pm kE \mp ii\,\mathbf{i}p_x) (\mp kE \mp ii\,\mathbf{i}p_z)$$

and two combinations of

$$(\pm kE \mp ii\,\mathbf{i}p_x) (\mp kE \mp ii\,\mathbf{i}p_x) \quad (\pm kE \mp ii\,\mathbf{j}p_y) (\mp kE \mp ii\,\mathbf{j}p_y)$$
$$(\pm kE \mp ii\,\mathbf{k}p_z) (\mp kE \mp ii\,\mathbf{k}p_z)$$

These structures are, of course, identical to an equivalent set in which both brackets undergo a complete sign reversal:

$$(\mp kE \pm ii\,\mathbf{i}p_x) (\pm kE \pm ii\,\mathbf{i}p_y) \text{ or } (\pm kE \pm ii\,\mathbf{i}p_y) (\mp kE \pm ii\,\mathbf{i}p_x), \text{ etc.}$$

'Colour' transitions can be seen as involving either an exchange of the components of **p** between the individual quarks, or as a relative switching of quark positions, so that the colours either move with the respective $p_x$, $p_y$, $p_z$ components, or switch with them. In either model the effect is the same, and a sign reversal in **p** is an additional necessary result. One method of picturing the exact symmetry presented in (16) is to imagine an automatic mechanism of transfer between the phases. And, since the *E* and **p** terms in the state vector really represent time and space derivatives, we can replace these with the covariant derivatives needed for invariance under a local *SU*(3) gauge transformation. A significant aspect of this *SU*(3) symmetry or *strong interaction* is that, because it depends entirely on the nilpotency of the component state vectors, it is entirely nonlocal. That is, the exchange of momentum **p** involved is entirely independent of any spatial position of the 3 components of the baryon. We can suppose, therefore, that the rate of change of momentum (or 'force') is constant with respect to spatial positioning or separation. A force that is constant with separation is equivalent to a potential which is linear with distance, exactly as is required for the conventional strong interaction.

It is significant that the symmetry evident in (16) requires equivalent status for the +**p** and –**p** states associated with positive energy. In other words, it requires the simultaneous existence of fermionic states of both negative and positive helicity, and so determines that the proton or any other state with baryonic structure must have finite (positive) mass. In principle, this immediately solves the mass gap problem. At the same time, the requirement of unbroken gauge invariance, which is a consequence of the vector nature of **p**, requires that the mediators must be massless, and so spin 1.

**10 Parities of bosons and baryons**

A good test of the validity of the state vectors proposed in the preceding sections for bosons and baryons is to see if they reproduce the known parities for these systems in their ground states. Defining the parity transformation on $\psi$ as $i \psi i$, and assuming that we have defined the correct wavefunctions bosons and baryons, we can now investigate if these give the correct intrinsic parities in the ground state. Applying the transformation to a scalar boson (spin 0), we obtain:

$$\boldsymbol{i}(\pm \boldsymbol{k}E \pm i\boldsymbol{i}\,\mathbf{p} + i\boldsymbol{j}\,m)(\mp \boldsymbol{k}E \mp i\boldsymbol{i}\,\mathbf{p} + i\boldsymbol{j}\,m)\boldsymbol{i} = -\boldsymbol{i}(\pm \boldsymbol{k}E \pm i\boldsymbol{i}\,\mathbf{p} + i\boldsymbol{j}\,m)\boldsymbol{i}\boldsymbol{i}(\mp \boldsymbol{k}E \mp i\boldsymbol{i}\,\mathbf{p} + i\boldsymbol{j}\,m)\boldsymbol{i}$$

$$= -(\pm \boldsymbol{k}E \mp i\boldsymbol{i}\,\mathbf{p} + i\boldsymbol{j}\,m)(\mp \boldsymbol{k}E \pm i\boldsymbol{i}\,\mathbf{p} + i\boldsymbol{j}\,m)$$

The total transformed wavefunction $i\psi i$ thus becomes $-\psi$, indicating that the original wavefunction had negative parity. For the vector meson (spin 1), the result is the same, because:

$$\boldsymbol{i}(\pm \boldsymbol{k}E \pm i\boldsymbol{i}\,\mathbf{p} + i\boldsymbol{j}\,m)(\mp \boldsymbol{k}E \pm i\boldsymbol{i}\,\mathbf{p} + i\boldsymbol{j}\,m)\boldsymbol{i} = -\boldsymbol{i}(\pm \boldsymbol{k}E \pm i\boldsymbol{i}\,\mathbf{p} + i\boldsymbol{j}\,m)\boldsymbol{i}\boldsymbol{i}(\mp \boldsymbol{k}E \pm i\boldsymbol{i}\,\mathbf{p} + i\boldsymbol{j}\,m)\boldsymbol{i}$$

$$= -(\pm \boldsymbol{k}E \mp i\boldsymbol{i}\,\mathbf{p} + i\boldsymbol{j}\,m)\boldsymbol{i}\boldsymbol{i}(\mp \boldsymbol{k}E \mp i\boldsymbol{i}\,\mathbf{p} + i\boldsymbol{j}\,m)\boldsymbol{i}$$

Trying the same operation on a baryon, we can take one of the terms, say,

$$(\boldsymbol{k}E + i\boldsymbol{j}\,m)(\boldsymbol{k}E + i\boldsymbol{j}\,m)(\boldsymbol{k}E + i\boldsymbol{i}\,\mathbf{p} + i\boldsymbol{j}\,m),$$

and apply a parity transformation, to give:

$$\boldsymbol{i}\,(\boldsymbol{k}E + i\boldsymbol{j}\,m)(\boldsymbol{k}E + i\boldsymbol{j}\,m)(\boldsymbol{k}E + i\boldsymbol{i}\,\mathbf{p} + i\boldsymbol{j}\,m)\,\boldsymbol{i}\,.$$

This time, we can write it in the form:

$$\boldsymbol{i}\,(\boldsymbol{k}E + i\boldsymbol{j}\,m)\,\boldsymbol{i}\,\boldsymbol{i}\,(\boldsymbol{k}E + i\boldsymbol{j}\,m)\,\boldsymbol{i}\,\boldsymbol{i}\,(\boldsymbol{k}E + i\boldsymbol{i}\,\mathbf{p} + i\boldsymbol{j}\,m)\,\boldsymbol{i}$$
$$= (\boldsymbol{k}E + i\boldsymbol{j}\,m)(\boldsymbol{k}E + i\boldsymbol{j}\,m)(\boldsymbol{k}E - i\boldsymbol{i}\,\mathbf{p} + i\boldsymbol{j}\,m)\,.$$

Taking the result over all the terms (three with **p**, and three with – **p**), we obtain:

$$(k E + i\boldsymbol{j}\, m)\ (k E + i\boldsymbol{j}\, m)\ (k E - i\boldsymbol{i}\, \mathbf{p} + i\boldsymbol{j}\, m)$$
$$(k E + i\boldsymbol{j}\, m)\ (k E + i\boldsymbol{i}\, \mathbf{p} + i\boldsymbol{j}\, m)\ (k E + i\boldsymbol{j}\, m)$$
$$(k E + i\boldsymbol{j}\, m)\ (k E - i\boldsymbol{i}\, \mathbf{p} + i\boldsymbol{j}\, m)\ (k E + i\boldsymbol{j}\, m)$$
$$(k E + i\boldsymbol{j}\, m)\ (k E + i\boldsymbol{j}\, m)\ (k E + i\boldsymbol{i}\, \mathbf{p} + i\boldsymbol{j}\, m)$$
$$(k E - i\boldsymbol{i}\, \mathbf{p} + i\boldsymbol{j}\, m)\ (k E + i\boldsymbol{j}\, m)\ (k E + i\boldsymbol{j}\, m)$$
$$(k E + i\boldsymbol{i}\, \mathbf{p} + i\boldsymbol{j}\, m)\ (k E + i\boldsymbol{j}\, m)\ (k E + i\boldsymbol{j}\, m)\ ,$$

and

$$\boldsymbol{i}\,\psi\,\boldsymbol{i} = \psi\,.$$

So a baryon wavefunction, as defined, would have positive parity. Such calculations, of course, apply to the ground state values only, because if extra angular momentum terms are added, then extra terms must be supplied to the state vectors, with the sign of parity reversing for each additional term.

## 11 Vacuum

A fermion creation operator $(kE + i\boldsymbol{i}\,\mathbf{p} + i\boldsymbol{j}\,m)$ is unaffected if postmultiplied by $\boldsymbol{k}\,(kE + i\boldsymbol{i}\,\mathbf{p} + i\boldsymbol{j}\,m)$ (if we assume that scalar factors are removed by normalisation). That is,

$(kE + i\boldsymbol{i}\,\mathbf{p} + i\boldsymbol{j}\,m) = (kE + i\boldsymbol{i}\,\mathbf{p} + i\boldsymbol{j}\,m)\,\boldsymbol{k}\,(kE + i\boldsymbol{i}\,\mathbf{p} + i\boldsymbol{j}\,m)\,\boldsymbol{k}\,(kE + i\boldsymbol{i}\,\mathbf{p} + i\boldsymbol{j}\,m)\,\boldsymbol{k}\,(kE + i\boldsymbol{i}\,\mathbf{p} + i\boldsymbol{j}\,m) \ldots$

The same applies if the operator is postmultiplied by $\boldsymbol{i}\,(kE + i\boldsymbol{i}\,\mathbf{p} + i\boldsymbol{j}\,m)$ or $\boldsymbol{j}\,(kE + i\boldsymbol{i}\,\mathbf{p} + i\boldsymbol{j}\,m)$. In effect, $\boldsymbol{k}\,(kE + i\boldsymbol{i}\,\mathbf{p} + i\boldsymbol{j}\,m)$, $\boldsymbol{i}\,(kE + i\boldsymbol{i}\,\mathbf{p} + i\boldsymbol{j}\,m)$ and $\boldsymbol{j}\,(kE + i\boldsymbol{i}\,\mathbf{p} + i\boldsymbol{j}\,m)$ act as vacuum operators, leaving the fermion state unchanged. However,

$(kE + i\boldsymbol{i}\,\mathbf{p} + i\boldsymbol{j}\,m) = (kE + i\boldsymbol{i}\,\mathbf{p} + i\boldsymbol{j}\,m)\,\boldsymbol{k}\,(kE + i\boldsymbol{i}\,\mathbf{p} + i\boldsymbol{j}\,m)\,\boldsymbol{k}\,(kE + i\boldsymbol{i}\,\mathbf{p} + i\boldsymbol{j}\,m)\,\boldsymbol{k}\,(kE + i\boldsymbol{i}\,\mathbf{p} + i\boldsymbol{j}\,m) \ldots$

can also be written as

$(kE + i\boldsymbol{i}\,\mathbf{p} + i\boldsymbol{j}\,m) = (kE + i\boldsymbol{i}\,\mathbf{p} + i\boldsymbol{j}\,m)\,(kE - i\boldsymbol{i}\,\mathbf{p} + i\boldsymbol{j}\,m)\,(kE + i\boldsymbol{i}\,\mathbf{p} + i\boldsymbol{j}\,m)\,(kE - i\boldsymbol{i}\,\mathbf{p} + i\boldsymbol{j}\,m) \ldots$

with alternate states implying antifermion creation; or with the whole operation implying alternate creations of fermion and boson.

Physically, we could suppose that the fermion sees in the vacuum produced by the operator $\boldsymbol{k}$ its 'image' or virtual antistate, producing a kind of virtual bosonic combination, and leading to an infinite alternating series of virtual fermions and bosons. Taking the fermion state as a whole, this links up with the idea that the supersymmetry operator $Q$ and its Hermitian conjugate $Q^\dagger$ are simply the respective fermion and antifermion operators, $(\pm k E \pm i i \mathbf{p} + i j m)$ and $(\mp k E \pm i i \mathbf{p} + i j m)$. In the context of renormalization, with this conception of vacuum, we could see an infinite succession of boson and fermion loops cancelling each other, without needing to generate a new set of supersymmetric partners. The bosons and fermions become their own supersymmetric partners.

In addition, the three vacuum coefficients, *k*, *i* and *j*, can be seen as originating in (or being responsible for) the concept of discrete (point-like) charge.

| | | |
|---|---|---|
| *k* (*kE* + *ii* **p** + *ij m*) | weak vacuum | fermion creation |
| *i* (*kE* + *ii* **p** + *ij m*) | strong vacuum | gluon plasma |
| *j* (*kE* + *ii* **p** + *ij m*) | electric vacuum | *SU*(2) |

In this interpretation, the charges act as a discrete partitioning of the continuous vacuum responsible for zero-point energy, with the separate conservation laws for weak, strong and electric charges implying that the three discrete partitions are entirely independent of each other. And the full fermion spinor (± *kE* ± *ii* **p** + *ij m*) can be seen as being equivalent to fermion creation plus three vacuum 'reflections', corresponding to the charge states:

| | |
|---|---|
| (*kE* + *ii* **p** + *ij m*) | fermion creation |
| (*kE* – *ii* **p** + *ij m*) | strong reflection |
| (– *kE* + *ii* **p** + *ij m*) | weak reflection |
| (– *kE* – *ii* **p** + *ij m*) | electric reflection |

Significantly, the 'weak' reflection is equivalent to a simultaneous switch from fermion to antifermion and (by preserving the sign of **p**) from one helicity state to another. The 'electric' reflection, on the other hand is a full charge conjugation (with no change in helicity), while the 'strong' reflection is a spin reversal.

It appears that the nilpotent state vector incorporates real and virtual components in the same way as mass and charge. *Zitterbewegung* is a switching between them. This is why state vectors are supersymmetric. It is a quantum equivalent of action plus virtual reaction. To put it in another way, the 'reflection' of a fermionic creation in a charged 'mirror' is equivalent to defining the rest of the universe for that creation, just as Newton's classical process of action and reaction is really between a body and the rest of the universe, rather than between two isolated bodies. Because we can only define a fermion by also defining the rest of the universe, the fermion itself is only half of the picture. This is what we mean by saying that a fermion has half-integral spin. The fermion state is incomplete without its vacuum (and, indeed, supersymmetric) partner; they are analogous to the action and reaction sides of a steady-state potential energy equation, with the fermion state alone represented by kinetic energy; and it is even possible to apply a classical kinetic energy equation for magnetic moment in a magnetic field to produce the ½-integral value of spin. In addition, fermions not only carry with them the virtual vacuum partners which describe their interactions with the rest of the universe, but they are also able, in appropriate circumstances, to realise them (singly) as real (mass-shell) partners, either through the creation of real boson or boson-like states or, in less compactified form, through applications of the Berry phase.

## 12 The self-energy of a free fermion

If supersymmetry is exact in the formalism proposed, then a free fermion in vacuum will produce its own loop cancellations and its energy will acquire a finite value without renormalization. Free fermion plus boson loops should cancel. That this is indeed the case can be shown by performing a basic perturbation calculation for first order coupling, and showing that it leads to zero in the case of a free fermion. Suppose we have a fermion acted on by the electromagnetic potentials $\phi$, $\mathbf{A}$. Then

$$\left( \pm k \frac{\partial}{\partial t} \pm ii\boldsymbol{\sigma}.\nabla + ijm \right)\psi = -e\left( \pm ik\phi \mp i\boldsymbol{\sigma}.\mathbf{A} \right)\psi$$

We now apply a perturbation expansion to $\psi$, so that

$$\psi = \psi_0 + \psi_1 + \psi_2 + \ldots,$$

where $\psi_0 = (kE + ii\boldsymbol{\sigma}.\mathbf{p} + ijm)\, e^{-i(Et - \mathbf{p}.\mathbf{r})}$ is the solution of the unperturbed equation:

$$\left( \pm k \frac{\partial}{\partial t} \pm ii\boldsymbol{\sigma}.\nabla + ijm \right)\psi = 0$$

and represents zeroth-order coupling, or a free electron of momentum $\mathbf{p}$.

Using the perturbation expansion, we can write

$$\left( \pm k \frac{\partial}{\partial t} \pm ii\boldsymbol{\sigma}.\nabla + ijm \right)(\psi_0 + \psi_1 + \psi_2 + \ldots) = -e\left( \pm ik\phi \mp i\boldsymbol{\sigma}.\mathbf{A} \right)(\psi_0 + \psi_1 + \psi_2 + \ldots),$$

leading to the series

$$\left(\pm k\frac{\partial}{\partial t}\pm ii\boldsymbol{\sigma}.\nabla+ijm\right)\psi_0=0$$

$$\left(\pm k\frac{\partial}{\partial t}\pm ii\boldsymbol{\sigma}.\nabla+ijm\right)\psi_1=-e(\pm ik\phi\mp i\boldsymbol{\sigma}.\mathbf{A})\psi_0,$$

$$\left(\pm k\frac{\partial}{\partial t}\pm ii\boldsymbol{\sigma}.\nabla+ijm\right)\psi_2=-e(\pm ik\phi\mp i\boldsymbol{\sigma}.\mathbf{A})\psi_1 \ldots$$

Expanding ($i\,k\phi - i\,\boldsymbol{\sigma}.\mathbf{A}$) as a Fourier series, and summing over **k**, we obtain

$$(i\,k\phi - i\,\boldsymbol{\sigma}.\mathbf{A}) = \Sigma\,(i\,k\phi(\mathbf{k}) - i\,\boldsymbol{\sigma}.\mathbf{A}(\mathbf{k}))\,e^{i\mathbf{k}.\mathbf{r}},$$

so that

$$\left(\pm k\frac{\partial}{\partial t}\pm ii\boldsymbol{\sigma}.\nabla+ijm\right)\psi_1=-e\sum(\pm ik\phi(\mathbf{k})\mp i\boldsymbol{\sigma}.\mathbf{A}(\mathbf{k}))e^{i\mathbf{k}.\mathbf{r}}\psi_0$$

$$=-e\sum(\pm ik\phi(\mathbf{k})\mp i\boldsymbol{\sigma}.\mathbf{A}(\mathbf{k}))e^{i\mathbf{k}.\mathbf{r}}(\pm kE\pm ii\boldsymbol{\sigma}.\mathbf{p}+ijm)e^{-i(Et-\mathbf{p}.\mathbf{r})}$$

$$=-e\sum(\pm ik\phi(\mathbf{k})\mp i\boldsymbol{\sigma}.\mathbf{A}(\mathbf{k}))(\pm kE\pm ii\boldsymbol{\sigma}.\mathbf{p}+ijm)e^{-i(Et-(\mathbf{p}+\mathbf{k}).\mathbf{r})}$$

Suppose we expand $\psi_1$ as

$$\psi_1=\sum v_1(E,\mathbf{p}+\mathbf{k})e^{-i(Et-(\mathbf{p}+\mathbf{k}).\mathbf{r})}.$$

Then

$$\sum\left(\pm k\frac{\partial}{\partial t}\pm ii\boldsymbol{\sigma}.\nabla+ijm\right)v_1(E,\mathbf{p}+\mathbf{k})e^{-i(Et-(\mathbf{p}+\mathbf{k}).\mathbf{r})}$$

$$=-e\sum(\pm ik\phi(\mathbf{k})\mp i\boldsymbol{\sigma}.\mathbf{A}(\mathbf{k}))(\pm kE\pm ii\boldsymbol{\sigma}.\mathbf{p}+ijm)e^{-i(Et-(\mathbf{p}+\mathbf{k}).\mathbf{r})}$$

and

$$\sum(\pm kE\pm ii\boldsymbol{\sigma}.(\mathbf{p}+\mathbf{k})+ijm)v_1(E,\mathbf{p}+\mathbf{k})e^{-i(Et-(\mathbf{p}+\mathbf{k}).\mathbf{r})}$$
$$=-e\sum(\pm ik\phi(\mathbf{k})\mp i\boldsymbol{\sigma}.\mathbf{A}(\mathbf{k}))(\pm kE\pm ii\boldsymbol{\sigma}.\mathbf{p}+ijm)e^{-i(Et-(\mathbf{p}+\mathbf{k}).\mathbf{r})}$$

and, equating individual terms,

$(\pm kE \pm ii\boldsymbol{\sigma}.(\mathbf{p}+\mathbf{k}) + ijm)\,v_1(E, \mathbf{p}+\mathbf{k}) = -e\,(\pm i\,k\phi(\mathbf{k}) \mp i\,\boldsymbol{\sigma}.\mathbf{A}(\mathbf{k}))\,(\pm kE \pm ii\boldsymbol{\sigma}.\mathbf{p} + ijm)$

We can write this in the form

$v_1(E, \mathbf{p}+\mathbf{k}) = -e[\pm kE \pm ii\,\boldsymbol{\sigma}.(\mathbf{p}+\mathbf{k}) + ijm]^{-1}\,(\pm i\,k\phi(\mathbf{k}) \mp i\,\boldsymbol{\sigma}.\mathbf{A}(\mathbf{k}))(\pm kE \pm ii\boldsymbol{\sigma}.\mathbf{p} + ijm)$

which means that

$$\psi_1=-e\sum[\pm kE\pm ii\boldsymbol{\sigma}.(\mathbf{p}+\mathbf{k})+ijm]^{-1}(\pm ik\phi(\mathbf{k})\mp i\boldsymbol{\sigma}.\mathbf{A}(\mathbf{k}))(\pm kE\pm ii\boldsymbol{\sigma}.\mathbf{p}+ijm)e^{-i(Et-(\mathbf{p}+\mathbf{k}).\mathbf{r})}$$

This is the wavefunction for first-order coupling, with an electron (for example) absorbing or emitting a photon of momentum **k**.

If we observe the process in the rest frame of the electron and eliminate any external source of potential, then **k** = 0, and ($ik\phi - i\boldsymbol{\sigma}.\mathbf{A}$) reduces to the static value, $ik\phi$. In this case, $\psi_1$ becomes

$$\psi_1 = -e[\pm k E \pm ii\boldsymbol{\sigma}.\mathbf{p} + ijm]^{-1}(\pm ik\phi)(\pm k E \pm ii\boldsymbol{\sigma}.\mathbf{p} + ijm)e^{-i(Et-\mathbf{p}.\mathbf{r})}$$

as the summation is no longer strictly required for a single order of the pure self-interaction. Writing this as

$$\psi_1 = -e(\mp k E \pm ii\boldsymbol{\sigma}.\mathbf{p} + ijm)(\mp k E \pm ii\boldsymbol{\sigma}.\mathbf{p} + ijm)(\pm ik\phi)e^{-i(Et-\mathbf{p}.\mathbf{r})}$$

we see that $\psi_1 = 0$, for any fixed value of $\phi$. Clearly, this will also apply to higher orders of self-interaction In other words, a *non-interacting* fermion requires no renormalization as a result of its self-energy.

Of course, an interacting fermion would still require 'renormalization', and this can be carried out using the usual methods of QED, QCD and QFD (though with a cut-off energy that emerges naturally from the gravitational-inertial event horizon as the Planck mass).[6] However, what this means is that a 'charge' actually acquires its intrinsic value or coupling constant from the 'vacuum', that is, from its interaction with all other charges of the same kind, exactly as we would expect from the definition of interactions, in nilpotent theory, as arising purely from vacuum 'reflections'. The so-called renormalization process would then be one of *scaling*, the scale not being fixed until the interaction with the 'rest of the universe' ('vacuum') was taken into account. It would be an exact expression of nonlocality, although preserving, of course, the 4-vector connection between space and time and energy and momentum characteristic of these interactions.

**13 BRST quantization**

The Dirac nilpotent operator, being automatically second quantized, already incorporates a full quantum field representation. More conventional approaches to field quantization, however, can be used to demonstrate the relation between charge and energy operators, which the nilpotent formalism requires. Nilpotent operators of a special kind are, in fact, already used in standard quantum field theory, and it will be instructive to make a direct link between these and terms of the form ($\pm kE \pm ii\mathbf{p} + ijm$), considered as both energy and charge operators. In the standard theory, field quantization requires gauge fixing before propagators can be constructed. The canonical quantization of the electromagnetic field uses Coulomb gauge, but this means that Lorentz invariance must be broken. The path integral approach allows us to use any gauge, and so maintain Lorentz invariance, but the problem now is the introduction of nonphysical or 'fictitious'

Fadeev-Popov ghost fields. A version used in string theory (BRST) eliminates the ghost fields by packaging all the information into a single operator, applied to the Lagrangian. Significantly, the BRST operator ($\delta_{BRST}$) is a nilpotent. This operator can be used to construct a Noether current ($J_\mu$), corresponding to a nilpotent BRST conserved fermionic charge ($Q_{BRST}$). The condition for defining a physical state then becomes

$$Q_{BRST} |\psi\rangle = 0 .$$

In the Dirac nilpotent formulation, ($\pm k E \pm i\mathbf{i}\mathbf{p} + i\mathbf{j}m$), which applies only to physical (mass shell) states, is already second quantized, and a nilpotent operator of the form $\delta_{BRST}$. It is, also, a nilpotent *charge* operator of the form $Q_{BRST}$, but extended to incorporate weak and strong, as well as electromagnetic, charges. It is, finally, in its eigenvalue form, identical to $|\psi\rangle$. So the three possible meanings for the expression ($\pm k E \pm i\mathbf{i}\mathbf{p} + i\mathbf{j}m$) apply, respectively, to: $E$ and $\mathbf{p}$ interpreted as differential operators in time and space; $E$, $\mathbf{p}$ and $m$ as coefficients determining the nature of the charges specified by $k$, $\mathbf{i}$ and $\mathbf{j}$; and $E$ and $\mathbf{p}$ interpreted as eigenvalues of energy and momentum. The nilpotent Dirac operator thus supplies simultaneously all the characteristics which the separate BRST terms $\delta_{BRST}$, $Q_{BRST}$, and $|\psi\rangle$ require.

## 14 The infrared divergence

The comprehensive packaging and defragmentation in the nilpotent formalism also removes anomalies in the theory of propagators. Conventional calculations separate the positive and negative energy states, leading to divergences, which are avoided in ordinary physical circumstances (though not removed) only by using cumbersome and rather arbitrary procedures involving contour integrals. Thus, in the Feynman formalism, the electron propagator is given by the expression

$$S_F(p) = \frac{1}{\slashed{p} - m} = \frac{\slashed{p} + m}{p^2 - m^2} .$$

which means that there is a singularity or 'pole' ($p_0$) where $p^2 - m^2 = 0$, the 'pole' or 'infrared divergence' being the point at which electrons switch to positron states. It is assumed that, on either side of the pole we have positive energy states moving forwards in time, and negative energy states moving backwards in time, the terms ($\slashed{p} + m$) and ($-\slashed{p} + m$) being used to project out, respectively, the positive and negative energy states. The procedure of avoiding the singularity requires adding the infinitesimal term $i\varepsilon$ to $p^2 - m^2$, and taking a contour integral over the complex variable, to give the solution

$$S_F(x-x') = \int d^3p \frac{1}{(2\pi)^3} \frac{m}{2E} \left[ -i\theta(t-t') \sum_{r=1}^{2} \Psi(x)\overline{\Psi}(x') + i\theta(t'-t) \sum_{r=3}^{4} \Psi(x)\overline{\Psi}(x') \right],$$

with summations over the up and down spin states.

Writing the denominator of the propagator as a nilpotent, however, makes the addition of an $i\varepsilon$ term unnecessary, because there is now no infrared divergence and no pole, as the denominator of the propagator term can be made into a positive nonzero scalar. We write

$$S_F(p) = \frac{1}{\pm kE \pm ii\boldsymbol{\sigma}.\mathbf{p} + ijm},$$

and are free to choose our usual interpretation of the reciprocal of a nilpotent:

$$\frac{1}{\pm kE \pm ii\boldsymbol{\sigma}.\mathbf{p} + ijm} = \frac{\mp kE \pm ii\boldsymbol{\sigma}.\mathbf{p} + ijm}{(\pm kE \pm ii\boldsymbol{\sigma}.\mathbf{p} + ijm)(\mp kE \pm ii\boldsymbol{\sigma}.\mathbf{p} + ijm)} = \frac{\mp kE \pm ii\boldsymbol{\sigma}.\mathbf{p} + ijm}{E^2 + p^2 + m^2},$$

which is finite at all values. The integral is now simply

$$S_F(x-x') = \int d^3p \frac{1}{(2\pi)^3} \frac{m}{2E} \theta(t-t') \Psi(x)\overline{\Psi}(x'),$$

in which

$$\Psi(x) = (\pm kE \pm ii\boldsymbol{\sigma}.\mathbf{p} + ijm) \exp(ipx),$$

and the adjoint term becomes

$$\overline{\Psi}(x') = (\pm kE \mp ii\boldsymbol{\sigma}.\mathbf{p} - ijm)...) (ik) \exp(-ipx'),$$

with ($\pm kE \mp ii\boldsymbol{\sigma}.\mathbf{p} - ijm$) ($ik$) a column vector. The reason for this success is apparent. The nilpotent formulation is automatically second quantized and the negative energy states appear as components of the nilpotent wavefunction on the same basis as the positive energy states. No averaging over spin states or 'interpreting' $-E$ as a reversed time state is necessary; the 'reversed time' state occurs with the $t$ in the operator $\pm E = \pm i\partial/\partial t$, and there is no need to separate out the states on opposite sides of a singularity.

The three boson propagators can be defined by analogy. Spin 1:

$$\Delta_F(x-x') = \frac{1}{(\pm kE \pm ii\boldsymbol{\sigma}.\mathbf{p} + ijm)(\mp kE \pm ii\boldsymbol{\sigma}.\mathbf{p} + ijm)},$$

Spin 0:

$$\Delta_F(x-x') = \frac{1}{(\pm kE \pm ii\boldsymbol{\sigma}.\mathbf{p} + ijm)(\mp kE \mp ii\boldsymbol{\sigma}.\mathbf{p} + ijm)},$$

and Bose-Einstein condensate / Berry phase:

$$\Delta_F(x-x') = \frac{1}{(\pm kE \pm ii\boldsymbol{\sigma}.\mathbf{p} + ijm)(\pm kE \mp ii\boldsymbol{\sigma}.\mathbf{p} + ijm)}.$$

Where the spin 1 bosons are massless (as in QED), we will have expressions like:
$$\Delta_F(x-x') = \frac{1}{(\pm kE \pm ii\boldsymbol{\sigma}.\mathbf{p})(\mp kE \pm ii\boldsymbol{\sigma}.\mathbf{p})} .$$

We can also perform virtually the same contour integral as in the case of the fermion to produce
$$i\Delta_F(x-x') = \int d^3p \frac{1}{(2\pi)^3} \frac{1}{2\omega} \theta(t-t')\phi(x)\phi^*(x') ,$$

where $\omega$ takes the place of $E / m$. This time, of course, $\phi(x)$ and $\phi(x')$ are scalar wavefunctions. In our notation, they are each scalar products of the 4-component bra term ($\pm kE \pm ii\boldsymbol{\sigma}.\mathbf{p} + ijm$) and the 4-component ket term ($\mp kE \pm ii\boldsymbol{\sigma}.\mathbf{p} + ijm$), multiplied respectively by exponentials exp ($ipx$) and exp ($ipx'$), expressed in terms of the 4-vectors $p$, $x$ and $x'$. In the nilpotent formulation, $\phi(x)\phi^*(x')$ reduces to a product of a scalar term (which can be removed by normalization) and exp $ip(x - x')$. In general, in off-mass-shell conditions, poles in the propagator are a mathematical, rather than physical problem; but, in the specific case of massless bosons, conventional theory states that 'infared' divergencies occur when such bosons are emitted from an initial or final stage which is on the mass shell. Such divergencies, however, will not occur where there is no pole.

The significant aspect of this analysis is that it shows that one of the principal divergences in quantum electrodynamics is (as the procedure used to remove it would suggest) merely an artefact of the mathematical structure we have imposed, and not of a fundamentally physical nature. As with the 'infinite' self-energy of the non-interacting fermion, it is a classic case of the action of a 'redundancy barrier'. Its automatic removal in the nilpotent formalism is another indication of the method's power and general applicability, and of the defragmentation process which it involves.

## 15 The Casimir effect

Though the weak, strong and electric partitions of the vacuum are discrete, reflecting the discrete nature of the fermions which define it, the combined vacuum may be thought of as continuous, in the sense that the action of a fermionic creation operator on any discrete realisation of it would immediately produce zero, as in:

($\pm kE \pm ii\,\mathbf{p} + ij\,m$) 1 ($\pm kE \pm ii\,\mathbf{p} + ij\,m$) = 0 .

In effect, the meaning of Pauli exclusion is that the total vacuum cannot be discrete, but must be a virtual realisation of the entire range of possible fermionic states. (We may perhaps consider that the action of a fermionic state on this continuum is to produce a singularity, a zeroing at that point in phase space.) A continuous vacuum, with an infinite range of vibration modes of zero-point energy $½\hbar\omega$, produces the well-known Casimir force of attraction between uncharged metal plates of area $A$ and small separation $d$:
$$F = \frac{\pi hc}{480} \frac{A}{d^4} .$$

Because of the dependence on $1/d^4$, the Casimir force manifests itself over the range 1 μm as a dipole-dipole interaction, and of exactly the same kind as the Van der Waals force of cohesion between molecules. This interpretation is based on zero-point fluctuations in the space between the plates or molecules, but, as Peterson and Metzger point out,[7] it is equally possible to represent the interaction as zero-point fluctuations of the electrons in the metal surfaces. In this case it becomes the London dispersion interaction. A third picture (Hellmann-Feynman) sees the quantum charge clouds in the two plates, molecules or other objects becoming deformed as they approach, corresponding to a change in the expectation values of their charge distributions. In this case, the force is identical to that of chemical bonding due to the classical electrostatic force.

Peterson and Metzger show, in effect, that the Casimir force is not a distinct phenomenon, but an aspect of the classical electromagnetic interaction. They use it as a means of removing such unobservables as quantum fluctuations from the argument, but we can turn the argument round so that the ordinary electromagnetic force becomes a vacuum projection. An inverse fourth power Casimir effect between objects, which are electrically neutral globally but composed locally of electrostatic dipoles, would then require an inverse square force between the individual charged particles of which they are composed. And there is no reason why the electrostatic force should be special – merely measurable at longer range than the others. Related effects of aggregated matter, for example nuclear forces, would become Casimir-type manifestations of vacuum fluctuations, either fermionic or bosonic, as much as interactions between discrete charges specified by expectation values.

Describing the forces due to discrete charges (electric, strong, weak) as Casimir-type manifestations of the vacuum, relates directly to the respective use of the quaternion operators *j*, *i*, *k* both for these three charges, and for the operation of the respective electric, strong, weak vacua via *j*($\pm$ *ikE* $\pm$ *i***p** + *jm*), *i*($\pm$ *ikE* $\pm$ *i***p** + *jm*), *k*($\pm$ *ikE* $\pm$ *i***p** + *jm*). Because the operators are attached respectively to pseudoscalar *E*, vector **p** and scalar *m*, in the state vector, then their vacua will have different effects, and so the forces will behave differently. However, the key driving mechanism in all Casimir calculations is that they are the result of separating out *discrete* objects from a *continuous* background, and that they only have meaning in the context of object pairs. The creation of a discrete object pair at some finite separation generates a force because it creates a discrete space which is shielded from some of the modes of vacuum vibration outside this space. In principle, this allows us to consider all interactions between discrete charged objects, and even the values of the charged coupling constants, as resulting from the *existence* of the rest of the universe as a vacuum state, exactly in line with renormalization and Mach's principle for the parallel case of inertial mass.

It seems, then, that we can take the Casimir and related effects as the way in which the discrete charged vacua manifest themselves in relation to the continuous total vacuum background; they represent the partitioning of the vacuum through the three types of charge state (or singularity). Whether the charge states are occupied or not (that is, have unit or zero values) is established on the basis of relative phases between the components of the state vector. Charge occupancy then determines particle type and possible interactions. The vacuum, however, is the mechanism by which this becomes manifested; the creation of discrete units with non-zero occupation status creates the 'distortions' of vacuum, which we call interactions, in the same way as the presence of discrete sources creates the vacuum response or distortions of simply-connected space which we call the Aharonov-Bohm effect and the Berry phase.

In general terms, vacuum may be thought of as the driving mechanism for assembly / disassembly and self-organization within aggregated matter, and for such things as phase transitions, effectively through the weak charge which defines fermionic matter. The Casimir effect will be attractive for bosons because they are weak dipoles, but repulsive for fermions because they are weak monopoles. The difference in status between the 'real' and image terms in the Dirac 4-spinor for a free particle, even if the 'real' particle is actually a vacuum state, also means that the Casimir effect does not require a broken supersymmetry to be observed, because loop cancellation is only at the level of the 'image' terms.

The creation and annihilation of fermions is, of course, one kind of phase transition, and involves the creation and annihilation of units of weak, and other, charges. The assembly of states of matter at other levels is equally concerned with the effect of the weak charge. Most of the properties of gaseous and condensed matter relate to the harmonic oscillator behaviour of its components, while the dipolar Van der Waals force, which expresses in its most fundamental aspect the nature of the weak vacuum, plays a significant role in all material phases. In addition, the properties of the solid state are determined by the Pauli exclusion principle that invariably accompanies the presence of weak charge, while Bose-Einstein condensation is effectively the elimination of this charge and its dipolarity, through the property of weak charge conjugation violation. Another phase transition of the Van der Waals-type occurs with the creation of interbaryonic, or nuclear, matter through a remnant of the strong forces between quarks, and this can be seen, in at least in part, as a Bose-Einstein condensation.

**16 *SU*(3)**

As we have seen, the vector nature of the **p** term in the Dirac nilpotent state vector produces a natural *SU*(3) symmetry for the strong interaction, which is reflected in the possible phases of the baryon state:

$$(\mathbf{k}E \pm i\mathbf{i}\, p_x + i\mathbf{j}\, m)\, (\mathbf{k}E \pm i\mathbf{i}\, p_y + i\mathbf{j}\, m)\, (\mathbf{k}E \pm i\mathbf{i}\, p_z + i\mathbf{j}\, m)\ .$$

The *SU*(3) symmetry thus expresses the perfect gauge invariance between all the possible phases, that is, with **p** = ± $p_x$, **p** = ± $p_y$, and **p** = ± $p_z$. The same would, of course, apply for those bosons (such as pions), which are held together by the strong interaction. The massless spin 1 gluons, which act as carriers, would have state vectors of the form (*kE* ± *ii* $p_x$) (∓ *kE* ± *ii* $p_y$), etc. Conventionally, we express the *SU*(3) symmetry via a 4-vector covariant derivative, which takes the form:

$$\partial_\mu \to \partial_\mu + ig_s \frac{\lambda^\alpha}{2} A^{\alpha\mu}(x).$$

In terms of the component coordinates, this becomes:

$$ip_i = \partial_i \to \partial_i + ig_s \frac{\lambda^\alpha}{2} A^{\alpha i}(x)$$

$$E = i\partial_0 \to i\partial_0 - g_s \frac{\lambda^\alpha}{2} A^{\alpha 0}(x).$$

And these can be inserted into the differential form of the baryon state vector, to obtain possible phases:

$$\left(k\left(E - g_s \frac{\lambda^\alpha}{2} A^{\alpha 0}\right) \pm i\left(\partial_1 + ig_s \frac{\lambda^\alpha}{2} \mathbf{A}^\alpha\right) + ijm\right)\left(k\left(E - g_s \frac{\lambda^\alpha}{2} A^{\alpha 0}\right) + ijm\right)\left(k\left(E - g_s \frac{\lambda^\alpha}{2} A^{\alpha 0}\right) + ijm\right)$$

$$\left(k\left(E - g_s \frac{\lambda^\alpha}{2} A^{\alpha 0}\right) + ijm\right)\left(k\left(E - g_s \frac{\lambda^\alpha}{2} A^{\alpha 0}\right) \pm i\left(\partial_1 + ig_s \frac{\lambda^\alpha}{2} \mathbf{A}^\alpha\right) + ijm\right)\left(k\left(E - g_s \frac{\lambda^\alpha}{2} A^{\alpha 0}\right) + ijm\right)$$

$$\left(k\left(E - g_s \frac{\lambda^\alpha}{2} A^{\alpha 0}\right) + ijm\right)\left(k\left(E - g_s \frac{\lambda^\alpha}{2} A^{\alpha 0}\right) + ijm\right)\left(k\left(E - g_s \frac{\lambda^\alpha}{2} A^{\alpha 0}\right) \pm i\left(\partial_1 + ig_s \frac{\lambda^\alpha}{2} \mathbf{A}^\alpha\right) + ijm\right)$$

which are exactly parallel to the six forms incorporated in the conventional antisymmetric baryon wavefunction:

$$\psi \sim (BGR - BRG + GRB - GBR + RBG - RGB),$$

based on the three quark 'colours' (*R*, *G*, *B*).

Through their relation to momentum components, the three colours are directly mapped onto the three dimensions of space, with the same indistinguishability, which is why the structure of real baryons is effectively affine, with a multiplicity of fractal-like structures involving gluons and virtual baryons beyond the arrangement of the valence quarks. We see from these structures that the 'active' term 'transferred' in the strong interaction is the vector term ($ig_s \lambda^\alpha \mathbf{A}^\alpha / 2$), which, in each phase, becomes the instantaneous carrier of the 'colour' component of the interaction, or, equivalently, the 'strong charge' (*s*). The concept of 'transfer' is, of course, a way of realising the superposition of all six phases, though, as we have already specified, it can be conceived of in terms of a scalar potential which is linear with distance. The scalar part of $A^{\alpha\mu}(x)$, however, is not transferred, and we will see that this coincides with an additional Coulomb component (or inverse linear potential) which is needed to define the spherical symmetry appropriate to a point source. In effect, this scalar or 'passive' component (which is a universal aspect of all fundamental interactions) is equivalent to the magnitude or scalar value of the strong coupling constant or strong charges. It has been shown, on the basis of reasonable assumptions, that, at Grand Unification, only this component would remain for the strong, weak and electric interactions, and that its value would be equivalent to the one expected for a gravitational or inertial force, namely the Planck mass.[8]

**17 $SU(2)_L \times U(1)$**

A 'weak interaction' is a demonstration that intrinsic left-handedness is an identical phenomenon in all fermionic states, while intrinsic right-handedness is an identical phenomenon in all antifermionic states, irrespective of the composition of the fermion or antifermion. So, the intrinsic handedness is preserved irrespective of any 'transition' between one state and another. All fermionic states, therefore, seek to demonstrate the gauge invariance of one-handedness with respect to all other possible fermionic states with probabilities determined by the energy, momentum and mass terms involved. This is what is meant by a 'transition'. In any such transition, the anti-state to the state to be annihilated and the state which is to be created must exist as a spin 1 bosonic combination. Because of quark confinement, there can be no transition from free fermion to quark, or quark to free fermion – that is, there can be no pure weak transition in which a fermion acquires or loses a 'vector' character. However, fermion states with mass also carry a degree of right-handedness. A non-vector transition from left- to right-handedness, involving only fermionic states (not antifermionic), requires the vacuum which we have described as 'electric'. Only the electric vacuum carries a transition to right-handedness where the vector character is absent, and, to produce a pure transition from left- to right-handedness (and *vice versa*) without a change from fermion to antifermion requires an electroweak combination (***jk***, equivalent to ***i***):

    ($\pm$ ***k****E* $\pm$ ***ii*** **p** + ***ij*** *m*)    left-handed fermion
    ($\mp$ ***k****E* $\pm$ ***ii*** **p** + ***ij*** *m*)    weak transition to right-handed antifermion
    ($\pm$ ***k****E* $\mp$ ***ii*** **p** + ***ij*** *m*)    electric transition to right-handed fermion

Using the concept of electric 'charge' as indicating the presence of right-handedness, we may identify four possible transitions (taking the 'left-handed' / 'right-handed' transition to mean 'the acquisition of a greater degree of right-handedness'), and hence four possible intermediate bosonic states:
        Left-handed to left-handed
        Left-handed to right-handed
        Right-handed to left-handed
        Right-handed to right-handed

The left- / right-handed transition clearly has the nature of an $SU(2)_L$ symmetry, with the requirement of three generators, which are necessarily massive, to carry the right-handedness unrecognised by the interaction, and two of which carry electric 'charge' (+ and –), in addition to one which leaves the handedness unchanged. This leaves the fourth transition state or equivalent as an extra generator with a $U(1)$ symmetry. If we assume that massive generators are necessary for a 'weak interaction', and indicate its presence, we can assign the fourth generator to the pure electric interaction. Electric charge, however, is not the sole reason for the massiveness (and hence mixed handedness) of real fermionic states. So the absence of electric charge does not indicate that a weak generator must be massless. So, the two generators without electric charge are assumed mixed, the combination producing two new generators, one of which becomes massless and so carries the pure electric, rather than the weak interaction.

To write the $SU(2)_L$ directly into the nilpotent representation, we can consider a lepton, for example, to be a superposition of states of the form ($\pm$ *kE* $\pm$ *ii* **p** + *ij m*) and ($\pm$ *kE* $\mp$ *ii* **p** + *ij m*), of which only the first acts weakly, while the neutrino is more likely to show Majorana behaviour as a superposition of ($\pm$ *kE* $\pm$ *ii* **p** + *ij m*) and ($\mp$ *kE* $\pm$ *ii* **p** + *ij m*). Baryons might be constructed in such a way that *each strong phase* is a superposition similar to that for the lepton. Further splitting of states into superpositions might be needed to fully incorporate the full range of fermionic particles within the three quark-lepton generations.

We can also relate the argument derived from the nilpotent structure to the conventional formalism for $SU(2)_L \times U(1)$, with $\mathbf{W}^\mu$ and $B^\mu$ as the respective 4-vector generators for $SU(3)$ and $U(1)$. Once again, we may write these down in the form of covariant derivatives. For left-handed states, we have:

$$\partial_\mu \to \partial_\mu + ig\frac{\boldsymbol{\tau}.\mathbf{W}^\mu}{2} - ig'\frac{B^\mu}{2},$$

and, for right-handed:

$$\partial_\mu \to \partial_\mu - ig'\frac{B^\mu}{2}.$$

The energy operator and the single well-defined component of spin angular momentum give us:

$$E = i\partial_0 \to i\partial_0 + ig'\frac{B^0}{2} + ig'\frac{B^3}{2},$$

and

$$ip_3 = \partial_3 \to \partial_3 + ig\frac{\boldsymbol{\tau}.\mathbf{W}^3}{2} + ig\frac{\boldsymbol{\tau}.\mathbf{W}^0}{2}.$$

So, we can write a vertex for a standard electroweak transition in the form:

$$(\pm k E \pm ii\mathbf{p} + ijm)\,(\mp k E \pm ii\mathbf{p} + ijm) =$$

$$\left(\pm k\left(\partial_0 + g'\frac{B^0}{2} + g'\frac{B^3}{2}\right) \pm i\left(\partial_3 + ig\frac{\boldsymbol{\tau}.\mathbf{W}^3}{2} + ig\frac{\boldsymbol{\tau}.\mathbf{W}^0}{2}\right) + ijm\right) \times$$

$$\left(\mp k\left(\partial_0 + g'\frac{B^0}{2} + g'\frac{B^3}{2}\right) \pm i\left(\partial_3 + ig\frac{\boldsymbol{\tau}.\mathbf{W}^3}{2} + ig\frac{\boldsymbol{\tau}.\mathbf{W}^0}{2}\right) + ijm\right)$$

With *m* determined from the combination of *E* and **p**, we can, by appropriate choice of the value of *m*, make these compatible, by additionally defining a combination of the coupling constants related to the $SU(2)_L$ and $U(1)$ symmetries, *g'* and *g*, which removes $B^3$ from *E* and $\mathbf{W}^0$ from **p**. It is, of course, significant here that it is $B^\mu$ which is characteristic of right-handed lepton states, and therefore associated with the production of mass. Writing these combinations as $\gamma^0$ (where $\gamma^0/2$ is equivalent to the electrostatic potential $\phi$) and $\mathbf{Z}^3$, and those of *g'* and *g*, as *e* and *w* (= *g*), we obtain:

$$(\pm k E \pm ii\mathbf{p} + ijm)\,(\mp k E \pm ii\mathbf{p} + ijm) =$$

$$\left(\pm k\left(\partial_0 + e\frac{\gamma^0}{2}\right) \pm i\left(\partial_3 + ig\frac{\boldsymbol{\tau}.\mathbf{Z}^3}{2}\right) + ijm\right)\left(\mp k\left(\partial_0 + e\frac{\gamma^0}{2}\right) \pm i\left(\partial_3 + ig\frac{\boldsymbol{\tau}.\mathbf{Z}^3}{2}\right) + ijm\right).$$

In many respects the weak and electric interactions are similar. The most significant similarity is the absence of any vector component. However, the pseudoscalar operator (incorporating the necessary mathematical pairing of +*i* and –*i*) determines that the weak interaction has an intrinsically dipolar source, but, because of the intrinsically one-handed nature of the interaction, this dipole does not accommodate two real charges, but one real charge and its virtual vacuum reflection. Another way of representing this is to say that the weak vacuum is filled for states negative energy (–*iE*), in much the same way as Dirac's original antimatter vacuum was filled (and for the same reason). It is because of this that the ground state of the 'rest of the universe' has no negative energy states, and that matter predominates over antimatter – the explanation requires physics, not cosmology. The weak vacuum requires an energy continuum in the same way as the conjugate variable time irreversibility requires continuity of time. It is precisely this continuity which makes possible both Pauli exclusion and its corollary, nonlocality. Pauli exclusion is a direct property of nilpotency, a mathematical condition which is only made possible by the presence of a pseudoscalar term. As we have shown, all these facts are consequences of the mathematical structure of the nilpotent Dirac operator.

Significantly, the *exchange* of electromagnetic charge, through $W^+$ or $W^-$, is not itself an electromagnetic interaction, but rather an indication of the weak interaction's indifference to the presence of the electromagnetic charge. A 'weak interaction', in principle, is a statement that all states of a particle with the same weak charge are equally probable, given the appropriate energy conditions, and that gauge invariance is maintained with respect to them. Weak bosons are massive because they act as carriers of the electromagnetic charge, whereas electromagnetic bosons (or photons) are massless because they do not. The quantitative value of the mass must be determined from the coupling of the weak charge to the asymmetric vacuum state which produces the violation of charge conjugation in the weak interaction. The weak interaction is also indifferent to the presence of the strong charge, and so cannot distinguish between quarks and leptons (hence, the intrinsic identity of purely lepton weak interactions with quark-lepton or quark-quark ones) and, in the case of quarks, it cannot tell the difference between a filled 'electromagnetic vacuum' (up quark) and an empty one (down quark). The weak interaction, in addition, is also indifferent to the sign of the weak charge, and responds (via the vacuum) only to the status of fermion or antifermion – hence, the Cabibbo-Kobayashi-Maskawa mixing.

**18 Charge structures**

The four components of the nilpotent Dirac spinor have been identified as the fermion and its three discrete vacuum 'reflections' under transitions which would have the characteristics of the weak, strong and electric interactions. The spinor can thus be considered as containing the full potentiality of what any fermionic state could be transformed into, and the weak, strong and electric interactions as the means of making this transfer. It is significant that the gravitational or inertial interaction is 'passive' in this respect, the vacuum reflection (expressible as $1\psi$ or scalar $\times \psi$) leading to the state itself. We can consider a fermion (with creation operator specified by the first term in the spinor) as having the potentiality to be switched by the appropriate interaction into any of the vacuum reflections that it carries with it, and that are specified by the quaternion operators labelled *k*, *i*, *j*, and that might be specified by the respective weak, strong and electric *charges*. In this respect, the action of the weak force becomes a change of fermion to antifermion, with a corresponding change in helicity; the action of the strong force becomes a change in helicity, without a change of fermionic status; while the action of the electric force becomes a change from fermion to antifermion, without a corresponding change in helicity.

There is, however, a significant distinction between the two types of fundamental fermion – quarks and leptons – in that only quarks incorporate the explicit vector behaviour of the momentum operator in their spinor state vectors. We can account for this distinction in terms of vector *phase*. So a baryon state vector might have a form such as

$$\begin{array}{l}\textit{inertial}\\ \textit{strong}\\ \textit{weak}\\ \textit{electric}\end{array} \begin{pmatrix} ikE \pm i\boldsymbol{\sigma}.\mathbf{p}_1 + jm \\ ikE \mp i\boldsymbol{\sigma}.\mathbf{p}_1 + jm \\ -ikE \pm i\boldsymbol{\sigma}.\mathbf{p}_1 + jm \\ -ikE \mp i\boldsymbol{\sigma}.\mathbf{p}_3 + jm \end{pmatrix} \begin{pmatrix} ikE \pm i\boldsymbol{\sigma}.\mathbf{p}_2 + jm \\ ikE \mp i\boldsymbol{\sigma}.\mathbf{p}_2 + jm \\ -ikE \pm i\boldsymbol{\sigma}.\mathbf{p}_3 + jm \\ -ikE \mp i\boldsymbol{\sigma}.\mathbf{p}_2 + jm \end{pmatrix} \begin{pmatrix} ikE \pm i\boldsymbol{\sigma}.\mathbf{p}_3 + jm \\ ikE \mp i\boldsymbol{\sigma}.\mathbf{p}_3 + jm \\ -ikE \pm i\boldsymbol{\sigma}.\mathbf{p}_2 + jm \\ -ikE \mp i\boldsymbol{\sigma}.\mathbf{p}_1 + jm \end{pmatrix}$$

or

$$\begin{array}{l}\textit{inertial}\\ \textit{strong}\\ \textit{weak}\\ \textit{electric}\end{array} \begin{pmatrix} ikE \pm i\boldsymbol{\sigma}.\mathbf{p}_1 + jm \\ ikE \mp i\boldsymbol{\sigma}.\mathbf{p}_1 + jm \\ -ikE \pm i\boldsymbol{\sigma}.\mathbf{p}_1 + jm \\ -ikE \mp i\boldsymbol{\sigma}.\mathbf{p}_3 + jm \end{pmatrix} \begin{pmatrix} ikE \mp i\boldsymbol{\sigma}.\mathbf{p}_2 + jm \\ ikE \pm i\boldsymbol{\sigma}.\mathbf{p}_2 + jm \\ -ikE \mp i\boldsymbol{\sigma}.\mathbf{p}_3 + jm \\ -ikE \pm i\boldsymbol{\sigma}.\mathbf{p}_2 + jm \end{pmatrix} \begin{pmatrix} ikE \pm i\boldsymbol{\sigma}.\mathbf{p}_3 + jm \\ ikE \mp i\boldsymbol{\sigma}.\mathbf{p}_3 + jm \\ -ikE \pm i\boldsymbol{\sigma}.\mathbf{p}_2 + jm \\ -ikE \mp i\boldsymbol{\sigma}.\mathbf{p}_1 + jm \end{pmatrix}$$

Here the charge reflections are all equally present, but have different *phases*, the phases being determined by that of the momentum operator, as determined by the strong interactions (which, in the baryonic system, would naturally be associated with that of the inertial or total particle state).

On the other hand, leptons, with no explicit vector phases, would have all components in phase at once in the successive manifestations of **p** as $\mathbf{p}_1$, $\mathbf{p}_2$ and $\mathbf{p}_3$, as in

$$\begin{array}{l}\textit{inertial}\\ \textit{strong}\\ \textit{weak}\\ \textit{electric}\end{array} \begin{pmatrix} ikE \pm i\boldsymbol{\sigma}.\mathbf{p}_1 + jm \\ ikE \mp i\boldsymbol{\sigma}.\mathbf{p}_1 + jm \\ -ikE \pm i\boldsymbol{\sigma}.\mathbf{p}_1 + jm \\ -ikE \mp i\boldsymbol{\sigma}.\mathbf{p}_1 + jm \end{pmatrix}$$

Such a consideration of phases (along with the symmetries known for each interaction) produces structures for all fundamental fermions indicating the charge activity in each phase for all vacuum reflections. The structures show that the expressions for the nilpotent Dirac spinors give direct information on charge activity, as well as on energy, momentum and rest mass (or space, time and proper time). Since there are five fundamental quantities or 'dimensions' associated with weak charge, strong charge (in three colours) and electric charge, along with five 'dimensions' associated with $E$, **p** and $m$, we can consider the nilpotent spinor as a mathematical object in a 10-dimensional phase space (with a conjugate real space) which provides all the basic requirements for a string theory without confinement to a specific model, along with an underlying algebraic structure with direct connections to such groups associated with string theory as $SO(32)$ and $E_8$ (and because of its origin in the eight fundamental units of multivariate vectors plus quaternions, an easy mapping to either an octonion or a twistor representation).[9,10] 'Dimensionality', of course, is definable in many ways; and the nilpotent operator can also be seen according to different criteria as 1-, 2-, 3-, 4-, 5-, 6-, 8- or even higher-dimensional, and a variety of different geometrical algebras can be used to create the 64 unit structures needed for the gamma matrices. The multiplicity of dimensionalities is provided by the fact the basic units contain two independent 3-dimensional systems. However, the 10-dimensional representation is exactly of the kind required by string theory. Also, the requirement for a perfect string theory is that self-duality in phase space determines vacuum selection. The nilpotent operator is self-dual, expressed in terms of phase space, and completely determines vacuum selection. Here, we have that requirement fulfilled exactly, without any need for an intermediate 'physical' model, and an embedding eleventh 'dimension' is provided by the Hilbert space within which the state vectors operate.

**19 The electric interaction: inverse linear potential**

Previously, we have speculated, on the basis of an extension of Noether's theorem, that the conservation of 'type of charge' (weak, strong or electric) corresponds exactly to the conservation of angular momentum.[11] As the behaviour related to these charges has now been located in the **p** or 'spin' term, which incorporates the angular momentum aspect of the fermionic state vector, we can now be more specific about the meaning of this connection. Essentially, charge is defined for a point source, with spherical symmetry. Spherical symmetry effectively determine angular momentum conservation, and has three fundamental aspects, namely the facts that the symmetry is independent of the length of the radius vector, of its direction (or choice of axes), and of the handedness of the rotation (left or right). We can see immediately that these correspond to the respective $U(1)$, $SU(3)$ and $SU(2)$ or $O(3)$ symmetries. We can also show, mathematically, that only three conditions exist in which a nilpotent state vector can maintain spherical symmetry.[12] All require the actions of scalar potentials: and these are inverse linear with distance, which corresponds to $U(1)$; direct linear with distance plus inverse linear, which corresponds to $SU(3)$; and *any other spherically symmetrical relation with distance* plus inverse linear. The last incorporates the special case of inverse third power plus inverse linear, which is characteristic of a dipole-dipole force, such as the pure weak interaction requires, and which provides a harmonic oscillator solution, exactly corresponding to the characteristic behaviour of the weak interaction as a creator and destroyer of fermion-antifermion pairs, or, in effect, weak dipoles.

The defragmented Dirac equation is remarkably easy to solve in these cases, as the energy, momentum and mass operators maintain their separate identities through all operations. This means that we don't need to break up the wavefunction into separate energy and momentum eigenfunctions to do calculations. The method is also completely general, and can be applied to any type of potential. In principle, all we need to do is to set up a differential operator with the appropriate field terms, and then find the function which will make its eigenvalue (taken over the four solutions equivalent to $\pm E$, $\pm \mathbf{p}$) a nilpotent. The method is exact and analytic and provides the full 'hydrogen atom' solution (involving hyperfine levels) in the case of the pure Coulomb interaction in just seven steps (though a few more will be added here for clarity). It could even be argued that it is more strictly correct than the conventional method, which assumes that amplitudes that vary with position can be treated as constant.

It will be convenient, in all these calculations, to use ordinary vectors, rather than multivariate vectors, so we can use the standard conversion of the $\nabla$ or $\boldsymbol{\sigma}.\nabla$ term to polar coordinates, with explicit introduction of fermionic spin. So we write:

$$\boldsymbol{\sigma}.\nabla = \left(\frac{\partial}{\partial r} + \frac{1}{r}\right) \pm i\frac{j + \frac{1}{2}}{r}.$$

We then set up an operator of the form:

$$\left(k(E + V(r)) + i\left(\frac{\partial}{\partial r} + \frac{1}{r} \pm i\frac{j + \frac{1}{2}}{r}\right) + ijm\right),$$

where $V(r)$ is the radially-dependent potential energy term. It will quickly become apparent that, unless $V(r)$ contains an expression of the form $A / r$, or $- A / r$, to compensate for those in the term beginning with $\boldsymbol{i}$, then no nilpotent solution can be found. We can regard this *Coulomb term* as the minimum requirement for spherical symmetry. It is, as we have previously stated, an expression of the magnitude of the charge, or the coupling constant. So, we begin with:

$$\left(k\left(E - \frac{A}{r}\right) + i\left(\frac{\partial}{\partial r} + \frac{1}{r} \pm i\frac{j + \frac{1}{2}}{r}\right) + ijm\right).$$

All we have to do is find the phase which will make the amplitude (or eigenvalue) nilpotent. So, we try the standard solution:

$$F = e^{-ar} r^{\gamma} \sum_{\nu = 0} a_{\nu} r^{\nu}.$$

Then

$$\frac{\partial F}{\partial r} = \left(-a + \frac{\gamma}{r} + \frac{\nu}{r} + \ldots\right) F,$$

where, for a bound state, $a$ is real and positive, and the amplitude produced by the differential operator then becomes

$$\left(\pm k\left(E - \frac{A}{r}\right) \pm i\left(-a + \frac{\gamma}{r} + \frac{\nu}{r} + \ldots \frac{1}{r} \pm i\frac{j + \frac{1}{2}}{r}\right) + ijm\right).$$

Squaring, and applying the nilpotency condition, we obtain:

$$4\left(E - \frac{A}{r}\right)^2 = -2\left(-a + \frac{\gamma}{r} + \frac{\nu}{r} + \ldots \frac{1}{r} + i\frac{j+\frac{1}{2}}{r}\right)^2 - 2\left(-a + \frac{\gamma}{r} + \frac{\nu}{r} + \ldots \frac{1}{r} - i\frac{j+\frac{1}{2}}{r}\right)^2 + 4m^2 .$$

Equating constant terms leads to

$$E^2 = -a^2 + m^2 ,$$
$$a = \sqrt{m^2 - E^2} .$$

Equating terms in $1/r^2$, with $\nu = 0$, we obtain:

$$\left(\frac{A}{r}\right)^2 = -\left(\frac{\gamma+1}{r}\right)^2 + \left(\frac{j+\frac{1}{2}}{r}\right)^2 ,$$

from which, excluding the negative root (as usual),

$$\gamma = -1 + \sqrt{(j+\frac{1}{2})^2 - A^2} .$$

Assuming the power series terminates at $n'$, and equating coefficients of $1/r$ for $\nu = n'$,

$$2EA = -2\sqrt{m^2 - E^2}\,(\gamma + 1 + n'),$$

the terms in $(j + \frac{1}{2})$ cancelling over the summation of the four multiplications, with two positive and two negative. From this we may derive

$$\frac{E}{m} = \frac{1}{\sqrt{1 + \dfrac{A^2}{(\gamma+1+n')^2}}} = \frac{1}{\sqrt{1 + \dfrac{A^2}{\left(\sqrt{(j+\frac{1}{2})^2 - A^2} + n'\right)^2}}} .$$

With $A = Ze^2$, we obtain the hyperfine or fine structure formula for a one-electron nuclear atom or ion:

$$\frac{E}{m} = \frac{1}{\sqrt{1 + \dfrac{(Ze^2)^2}{(\gamma+1+n')^2}}} = \frac{1}{\sqrt{1 + \dfrac{(Ze^2)^2}{\left(\sqrt{(j+\frac{1}{2})^2 - (Ze^2)^2} + n'\right)^2}}} .$$

and, with $Z = 1$, it becomes applicable to hydrogen.

## 20 The strong interaction: linear plus inverse linear potential

The inverse linear potential gives the scalar phase solution expected for a *U*(1) symmetry, such as we have associated with the electric interaction, where the charge is a pure scalar magnitude. We have already shown that a direct linear potential provides the requirements for the vector term needed for the strong interaction, though once again a Coulomb term is needed for magnitude and for spherical symmetry. No other special solution exists which gives nilpotency along with spherical symmetry. In principle, it is not significant whether the source of the strong field on a quark is an antiquark in a mesonic combination, or a strong centre of charge acting equally on the three quark components of a baryon. The solution will be structurally the same in each case, differing only in the specific values for numerical constants.

First of all we set up an operator of the form:

$$\left( k\left( E - \frac{A}{r} - Br \right) + i\left( \frac{\partial}{\partial r} + \frac{1}{r} \pm i\frac{j+½}{r} \right) + ijm \right).$$

We can now easily guess that the phase term required has the structure

$$exp(-ar - br^2) r^\gamma \sum_{v=0} a_v r^v \ ,$$

and consider the ground state (with $v = 0$) over the four Dirac solutions. Applying the differential operator and imposing the nilpotency condition, we obtain:

$$E^2 + 2AB + \frac{A^2}{r^2} + B^2 r^2 - \frac{2AE}{r} - 2BEr = m^2$$

$$-\left( a^2 + \frac{(\gamma+v+...+1)^2}{r^2} - \frac{(j+½)^2}{r^2} + 4b^2 r^2 + 4abr - 4b(\gamma+v+...+1) - \frac{2a}{r}(\gamma+v+...+1) \right)$$

The positive and negative $i(j + ½)$ terms again cancel out over the four solutions as they do in the case of the hydrogen atom, and, assuming a termination in the power series, we can equate:

(1) coefficients of $r^2$:     $B^2 = -4b^2$
(2) coefficients of $r$:      $-2BE = -4ab$
(3) coefficients of $1/r$:    $-2AE = 2a(\gamma+v+1)$
(4) coefficients of $1/r^2$:  $A^2 = -(\gamma+v+1)^2 + (j+½)^2$
(5) constant terms:           $E^2 + 2AB = -a^2 + 4b(\gamma+v+1) + m^2$

From the first three equations, we immediately obtain:
$$b = \pm \frac{iB}{2}$$
$$a = \mp iE$$
$$\gamma + \nu + 1 = \mp iA .$$
The case where $\nu = 0$ then requires a phase term
$$\psi = exp\left(\pm iEr \mp iBr^2/2\right) r^{\mp iqA - 1} .$$

The imaginary exponential terms in $\psi$ can be interpreted as representing asymptotic freedom, the exp ($\pm iEr$) being typical for a free fermion. The $r^{\gamma-1}$ term is also complex, and can be written as a phase, $\phi(r) = \exp(\mp iqA \ln(r))$, which varies less rapidly with $r$ than the rest of $\psi$. We can therefore write $\psi$ in the form
$$\psi = \frac{exp(kr + \phi(r))}{r} ,$$
where
$$k = \pm iE \mp iBr/2 .$$
Where $r$ is small (at high energies), the first term dominates, approximating to a free fermion solution (which can be interpreted as asymptotic freedom). When $r$ is large (at low energies) the second term dominates, bringing in the confining potential ($B$) (which can be interpreted as infrared slavery). Significantly, the phase term $\phi$ incorporates the Coulomb or scalar phase component. Reducing the potential $V(r)$ to the Coulomb term, which is what we suppose might happen effectively at short distances, produces a hydrogen-like spectral series, with exactly the same structure as in the previous section.

We can use the full and Coulomb-like solutions to investigate the transition point at which infrared slavery becomes effective. From the full solution, let
$$k = \pm iE \mp i\frac{Br}{2} = \frac{2\pi}{\lambda} = 0$$
at zero effective energy (or infrared slavery). Then
$$r = \frac{2E}{B}$$
If, from the Coulomb-like solution, we take the 'free-particle' transition energy as the mass of the state $m$, and assume that this mass is mostly dynamic (gluonic) in origin, then we find $Br = 2E$, suggesting a virial relationship, as would be expected with a linear potential.

The calculation in this section has been performed for a quark-antiquark pair. Extending it to a 3-quark system, we can expect that the magnitude of the 'active' coupling is essentially identical, while the 'passive' or Coulomb coupling is related by a virial factor, so that $A_{3Q} \approx A_{QQ}/2$ (in line with the theoretically-assumed value of $2\alpha_s/3$ for the magnitude of $A$). This would accord with the model of Takahashi *et al*, based on lattice gauge QCD,[13] in which these relations between coupling constants apply. A constant term in the potential, as incorporated by these authors, would have no effect on the structure of the phase term, merely displacing the value used for *E*.

The nilpotent structures may be used to suggest explanations of some aspects of the interaction of protons, as well as quarks. At intermediate distance, this interaction looks principally like the exchange of a single gluon, with minor adjustments needed to maintain the colour singlet state. For any given phase, each proton will always be effectively represented as though by a single fermionic state, say (***k**E*$_1$ ± *i**i* **p**$_1$ + *i**j* *m*$_1$) and (***k**E*$_2$ ± *i**i* **p**$_2$ + *i**j* *m*$_2$). Now, the strong interaction between two such states, each in a single (though unspecified) phase, can be represented by a momentum transfer, which is exactly of the form involved in gluon exchange between the component parts of the proton. This will also be still beyond the threshold at which the Coulombic or scalar part of the interaction dominates over the vector part.

However, the interaction between protons within a nuclear-type structure (separation 1 fm) is a saturated potential, of the form $r^n$, where $n \leq 2$, or polynomial combinations of that form. This is characteristic of a dipolar or multipolar force, and, of course, nuclear matter exists in an energy regime at which the interacting particle is a strong-dipolar massive pion involved in the Coulombic or scalar part of the strong force, rather than the massless gluon exchange involved in the vector part. In effect, the pion is created because nuclear matter has undergone a phase transition, in which the weak force also plays a part, for the pion is a weak, as well as a strong dipole. The weak force appears to be involved generally within phase transitions because it is *fundamentally* dipolar, in response to its origin within the pseudoscalar, or energy, term in the Dirac nilpotent state. By this, we mean that the weak interaction creates a dipole between the fermion state and the vacuum (if the fermion states has no real partner), leading to spin ½ and *zitterbewegung*, and the weak interaction, as a uniquely one-handed force, also uniquely has a dipole moment, which is manifested through the spin. It is this dipole moment which is responsible for the tendency of fermions to structure themselves as aggregated matter, finding real, rather than vacuum, partners among other fermions or fermion-like structures, and creating dipolarity and multipolarity within the other forces. A weak interaction between two fermionic sources always includes the vacuum partner as the other dipole component. Dipolar / multipolar states are associated with harmonic oscillator-type regimes, or creation and annihilation processes, and it is precisely such processes which are involved in the concept of phase transition. The pionic state is essentially a colour or strong singlet because of the necessity of making a bosonic state a weak singlet.

## 21 The weak interaction: polynomial plus inverse linear potential

Outside of the two special cases of inverse linear and linear plus inverse linear potentials, a combination of *any other* polynomial variation (or combination of polynomial variations) with *r* plus inverse linear, with spherical symmetry, yields exactly the same solution: the harmonic oscillator. This is exactly what we would expect for the weak interaction, whose potential is fundamentally inverse cube plus inverse linear, in the basic dipolar form, with higher inverse orders of *r* for multipolar cases. (There is an intriguing parallel with the classical case of planetary perihelion precession, a combination of spin and orbital motion which emerges from a potential of exactly the same form, also derived from multipolar sources.) Let us assume that the nilpotent operator can be written in the form

$$\left( k\left( E - \frac{A}{r} - Cr^n \right) + i\left( \frac{\partial}{\partial r} + \frac{1}{r} \pm i\frac{j + \tfrac{1}{2}}{r} \right) + ijm \right),$$

where *n* is an integer greater than 1 or less than –1.

As usual, we need to find the phase which will make the amplitude nilpotent. Polynomial potential terms which are multiples of $r^n$ require the incorporation into the exponential of terms which are multiples of $r^{n+1}$. So, extending our work on the hydrogen atom and the strong interaction, we may suppose that the phase is of the form:

$$F = \exp(-ar - br^{n+1}) r^\gamma \sum_{\nu=0} a_\nu r^\nu .$$

So

$$\frac{\partial F}{\partial r} = \left( -a + (n+1)br^n + \frac{\gamma}{r} + \frac{\nu}{r} + \ldots \frac{1}{r} \ldots \right) F .$$

Applying this and the nilpotency condition, with a termination in the series, we obtain

$$4\left( E - \frac{A}{r} - Cr^n \right)^2 = -2\left( -a + (n+1)br^n + \frac{\gamma}{r} + \frac{\nu}{r} + \frac{1}{r} + i\frac{j+\tfrac{1}{2}}{r} \right)^2$$

$$-2\left( -a + (n+1)br^n + \frac{\gamma}{r} + \frac{\nu}{r} + \frac{1}{r} - i\frac{j+\tfrac{1}{2}}{r} \right)^2$$

Equating constant terms, we find

$$a = \sqrt{m^2 - E^2} \qquad (17)$$

Equating terms in $r^{2n}$, with $\nu = 0$:

$$C^2 = -(n+1)^2 b^2$$

$$b = \pm \frac{iC}{(n+1)} .$$

Equating coefficients of *r*, where $\nu = 0$:

$$AC = -(n+1) b (1 + \gamma) ,$$
$$(1 + \gamma) = \pm iA .$$

Equating coefficients of $1/r^2$ and coefficients of $1/r$, for a power series terminating in $\nu = n'$, we obtain

$$A^2 = -(1 + \gamma + n')^2 + (j + \tfrac{1}{2})^2 \qquad (18)$$

and

$$-EA = a(1 + \gamma + n') . \qquad (19)$$

Combining (17), (18) and (19) produces:

$$\left(\frac{m^2 - E^2}{E^2}\right)(1+\gamma+n')^2 = -(1+\gamma+n')^2 + (j+½)^2$$

$$E = -\frac{m}{j+½}(\pm iA + n').$$

If, in the case of the weak interaction, we associate *A*, the phase term required for spherical symmetry, with the random directionality of the fermion spin, we may assign to it a half-unit value (± ½ *i*), and obtain a set of energy levels of the form expected in the simple harmonic oscillator:

$$E = -\frac{m}{j+½}(½ + n').$$

In this case, the phase term, with its unit value and imaginary coefficient, will be introduced when the spin component is added to the σ.∇ term in transforming from rectilinear to polar coordinates. There may be some significance in the fact that the potential function may be assumed complex in certain circumstances, for this is the exact condition needed for *CP* violation which occurs, uniquely, in the weak interaction.

## 22 Conclusion

Redundancy in relativistic quantum mechanics has been associated with the singularities produced when the momentum operator is treated mathematically in a way that makes it rotation asymmetric. This can be overcome by the use of an algebra which is fundamentally 3-dimensional. It then appears that the fragmentation of the momentum operator in the conventional representation is totally avoidable, leading to a much more coherent picture of the fermionic state. It is clear from this procedure that much of the mathematical apparatus associated with the Dirac equation is totally redundant. In the nilpotent version which emerges, there is only a one-line operator, which has the same form whether the fermion is free or interacting. There is no need, in principle, to define a wavefunction, as the phase and amplitude is determined uniquely by the operator. There is no need, in addition, to define either operator or wavefunction as a 4-component spinor, as none of the terms is totally independent of the others. All we need to specify is that any product of two quaternionic fermionic states will always result in a scalar value, this being the only contribution made by the '4-component' structure. In addition, there is only one phase, whether the system is fermionic, antifermionic or bosonic. Physically, of course, the quaternion labels, **k**, **i**, and **j**, which provide the additional 'solutions', have multiple functions, which now can be seen to be related in a fundamental way. They are, respectively, *T*, *P*, and *C* operators; and the generators of weak, strong, and electric vacua; they are also involved in the respective production of spin 1 bosons, Bose-Einstein condensates, and spin 0 bosons.

Many authors have applied geometrical and hypergeometrical algebras to the Dirac equation to make it easier to manipulate or more amenable to physical interpretation, with varying degrees of success; but there seems to be only one approach which gives a precise and exact solution to the fundamental problems of particle physics and relativistic quantum mechanics, and it requires all its elements in exactly the right place before it displays its full power. Ultimately, it would seem that the approach is also the simplest, its defragmentation of the Dirac equation making analytic calculation in significant examples relatively easy. Relativistic calculations turn out to be easier than nonrelativistic ones because all the elements are positioned in their correct places. Physical interpretations and explanations emerge purely from the nilpotent structure. Preliminary work on the less tractable problems posed by aggregated matter suggests that it might have just as significant an effect in those areas as well.